\documentclass{aa}  

\usepackage{graphicx}
\usepackage{txfonts}
\usepackage{lipsum}
\usepackage{subcaption}         % necessary for continued figures, example in section 3
\usepackage{lscape}             % to rotate a single page table, example in appendix.
\usepackage{placeins}           % useful with \FloatBarrier, to keep 
\newcommand{\caii}{\ion{Ca}{ii}}
\newcommand{\fei}{\ion{Fe}{i}}
\newcommand{\hi}{\ion{H}{i}}
\newcommand{\mgii}{\ion{Mg}{ii}}
\newcommand{\oi}{\ion{O}{i}}
\newcommand{\siiv}{\ion{Si}{iv}}
\newcommand{\cli}{\ion{Cl}{i}}
\newcommand{\kms}{km$\,$s$^{-1}$}
\newcommand{\Nmtwo}{N$\,$m$^{-2}$}
\newcommand{\nW}{nW/m$^2$/Hz/sr}
\newcommand{\pW}{pW/m$^2$/Hz/sr}

\begin{document}

%%%%%%%%%%%%%%%%%%%%%%%%%%%%%%%%%%%%%%%%
% if you use custom commands in your title,
% ensure to check your title when submitting!
%%%%%%%%%%%%%%%%%%%%%%%%%%%%%%%%%%%%%%%%
\title{Chromospheric dynamics and the O I 135.6~nm spectral line}
\subtitle{}

%%%%%%%%%%%%%%%%%%%%%%%%%%%%%%%%%%%%%%%%
% Please separate each author with the \and command
%
% Use the \corrauth to provide the corresponding
% author address. It will be automatically inserted as 
% footnote in the PDF output.
%
% Please DO NOT include ORCIDs next to author names.
% Instead, please provide an active address for each coauthor:
% it will be automatically extracted by EDPS editorial system, 
% and co-authors will be be able to authenticate their ORCID.
%
% Only authenticated ORCIDs will be taken into account.
% ORCIDs included here will be removed.
%%%%%%%%%%%%%%%%%%%%%%%%%%%%%%%%%%%%%%%%
\author{{Viggo H. Hansteen}\inst{1,2,3,4}
     \corrauth{vhansteen@seti.org}
\and
{Mats Carlsson}\inst{3,4}     \email{mats.carlsson@astro.uio.no}
\and
{Bart De Pontieu}\inst{1,3,4}
     \email{bdp@lmsal.com}
{Daniel N\'obrega-Siverio}\inst{5,6,3,4}
     \email{dnobrega@iac.es}
}

\institute{Lockheed Martin Solar \& Astrophysics Laboratory,
3251 Hanover St, Palo Alto, CA 94304, USA
\and
SETI Institute, 339 N Bernardo Ave Suite 200, Mountain View, CA 94043
\and
Rosseland Centre for Solar Physics, University of Oslo,  PO Box 1029 Blindern, N-0315 Oslo, Norway
\and
Institute of Theoretical Astrophysics, University of Oslo, PO Box 1029 Blindern, N-0315 Oslo, Norway
\and
Instituto de Astrof\'isica de Canarias, E-38205 La Laguna, Tenerife, Spain
\and
Universidad de La Laguna, Dept. Astrof\'isica, E-38206 La  Laguna, Tenerife, Spain}

\date{Received May 1, 2026}

% \abstract{}{}{}{}{}
% 5 {} token are mandatory
 
  \abstract
  % context heading (optional)
{The \oi\ 135.6~nm spectral line is formed in the chromosphere at the same heights as the \mgii~h\&k line cores are formed. As the \oi~line is optically thin, it represents a possibility for measuring the non-thermal velocities in this region without the complications added by optically thick radiative transfer. Numerical models have hitherto strained to reproduce \mgii\ core line widths, challenging current understanding of chromospheric energetics and dynamics. } 
  % aims heading (mandatory)
   {We aim to construct numerical models, varying physical and numerical parameters in order to asses which of these is most important in setting the \mgii\ core intensity and width.} 
  % methods heading (mandatory)
   {A set of numerical models of varying resolution, size, magnetic topology and strength are considered and used to synthesize \oi\ line emission and to investigate the constraints that observations of this line place on chromospheric dynamics and densities.}
  % results heading (mandatory)
   {We find that, for quiet Sun, while non-thermal motions undeniably provide a source of Doppler broadening and chromospheric mass loading, the average strength of the photospheric magnetic field is the most important parameter in setting the \mgii\ core width to values within 5~\kms\ of observed values. Furthermore, for plage, we identify non-equilibrium hydrogen ionization and three dimensional radiative transfer as important ingredients in understanding chromospheric diagnostics and deciphering chromospheric structure.}
  % conclusions heading (optional), leave it empty if necessary
   {}

 \maketitle
 \nolinenumbers

%%%%%%%%%%%%%%%%%%%%%%%%%%%%%%%%%%%%%%%%%%%%%%%%%%%%%%%%%%%%%%
\section{Introduction} \label{sec:introduction}

The upper chromosphere is the region of the solar atmosphere where the magnetic field first assumes a dominant role.  While the hydrodynamic forces control the lower chromosphere (up to 750~km or so above the photosphere), the $\beta \equiv p/(B^2/2\mu_0) = 1$ layer marks a transition. From this point outwards and into the interplanetary space, or at least to the Alfv\'en point some 10--20 $R_\odot$ from the solar surface, the magnetic field largely determines the 
energetics and dynamics of the lower heliosphere. The upper chromosphere is the recipient of and conduit through which all ``non-thermal'' or ``mechanical'' energy generated in the regions below passes, and in which dissipation of this energy can and will have important magnetic components. Thus we expect that, in this region, reconnection, Alfv\'enic wave dissipation, etc. will contribute significantly to the energy and force balance. In addition, we also expect that turbulent pressure gradients, acoustic wave dissipation, and similar will play important roles. 

The chromosphere is mainly diagnosed through a few strong optically thick absorption lines: the \caii~854.2~nm triplet line, \caii~H\&K, \mgii~h\&k, and \hi~H$_\alpha$ and Ly$_\alpha$. The cores of the latter lines are primarily formed in the upper $\beta < 1$ portion of the chromosphere. Since its launch in 2013, the Interface Region Imaging Spectrograph \citep[IRIS,][]{2014SoPh..289.2733D} has observed \mgii~line spectra for a plethora of solar scenes. The interpretation of optically thick spectral lines in terms of plasma conditions is possible but not straightforward. Fortunately, IRIS also observes spectral windows that cover other chromospheric and transition region lines, one of which is the unique chromospheric line \oi~135.6~nm line, which is formed under optically thin conditions \citep{2015ApJ...813...34L}. This line holds large potential to disentangle, as described below, the relative roles of magnetic and pressure driven mechanisms for heating the chromosphere.

Recent numerical models \citep{2023ApJ...944..131H,2024A&A...692A...6O}, computed using the 
Bifrost and MURaM codes, reproduce many aspects of the chromosphere, as evidenced by the \mgii\ line, but the synthetic spectra also show discrepancies, some of which are significant. This is especially apparent when considering the \mgii\ h\&k line cores which are sensitive to middle and upper chromospheric conditions. Comparisons of modeled and observed \mgii\ spectra show that modeled spectra often came out too faint, or too narrow, indicating a lack of opacity (density/temperature, i.e., heating), or alternately a lack of sufficient small scale motions in the chromosphere. The models described in \citet{2024A&A...692A...6O} point towards strong maximum-velocity differences or turbulent velocities in the chromosphere and lower atmosphere as being necessary to reproduce the observed line widths. On the other hand \citet{2023ApJ...944..131H} argued that flux emergence, perhaps arising from the magnetic field produced by a small scale dynamo, could be one of the main actors in determining the structure of the upper chromosphere. 

A possible discriminant between these views lies in observations of the \oi\ line and in particular its width, as described by \citet{2015ApJ...813...34L}. This line is optically thin, and hence observations of the line profile provide direct information about the physical state of the chromosphere in the region of the line's formation. In particular, the non-thermal broadening of the line can provide insight into the amplitude of any unresolved small-scale
motions. As reported by \citet{2015ApJ...813...34L} the \oi\ line is formed at the same heights as the \mgii\ core. As a result, observations should provide insight into the processes determining the core width of \mgii\ and thereby the main actor in upper chromospheric dynamics.

Following up on this work, \citet{2023ApJ...959...87C} show
that the non-thermal broadening of the \oi\ line is observed to be modest, with typical values of 5–10~\kms. Furthermore, the non-thermal broadening shows a modest but significant enhancement
above locations that are in between photospheric magnetic flux concentrations in plage, i.e., where the magnetic
field is likely to be more inclined with respect to the line of sight. These observations provide strict constraints on models of the chromosphere.

On the other hand, there are other possible effects determining the upper chromospheric structure and mass loading, and/or the resulting shape and intensity of the \mgii\ core, e.g., 3D
effects on the radiative transfer \citep{2020ApJ...901...32J}, the lack of sufficient numerical spatial resolution in the models computed, and/or effects that go beyond the standard MHD description of the chromospheric plasma, including, e.g., ambipolar diffusion
and non-equilibrium (NEQ) ionization of H and He.

In this paper we compute spectra of the \oi\ line for numerical models of various solar scenes: quiet Sun, plage, a coronal bright point (CBP), and an emerging flux region. The average photospheric field strength is varied in the quiet Sun models, as is the effect of changing the numerical resolution in the CBP model. In the plage models, we compare LTE hydrogen ionization with hydrogen populations computed using the non-equilibrium rate equations. We also study the differences incurred using the column by column RH1.5D radiative transfer code \citep{2015A&A...574A...3P,2001ApJ...557..389U} versus the Multi3D code \citep{Leenaarts_Carlsson:2009}.  

The main goal of this paper is to describe the formation of the \oi~line and in particular to study its $(1/e)$ width in the context of the numerical models mentioned above and compare this to what has been observed \citep[i.e.,][]{2023ApJ...959...87C}.  We will then study if there is any correlation between the width of the \oi~line and the core width of \mgii. The former gives a good estimate of the {\it most probable} velocity distribution at the formation height of the \mgii~line core. Thus, if such a correlation exists it would indicate that the (upper) chromosphere is dominated by small scale plasma dynamics and therefore that, e.g., the resolution at which a numerical simulation is run should play the most important role in setting the \mgii~width. On the other hand, if no such correlation exists we should consider whether other sources of broadening are dominant. For example, we will consider whether increasing the mean magnetic field strength is important, as it may change the chromospheric scale height. The topology of the magnetic field and its evolution may also change the chromospheric scale height, while more violent reconnection events in the chromosphere would increase the most probable velocity there. Furthermore, the inclusion of non-equilibrium hydrogen ionization will increase the electron number density $n_{\rm e}$ and hence the opacity in the regions where the \mgii~contribution function is large. This, in turn, can then broaden the core width and also increase the line intensity of the \oi-line. Finally, we will investigate whether using a fully three dimensional radiative solver (Multi3D) instead of computing \mgii-line profiles column by column (RH1.5D) can increase the core width. 

\section{Model description} \label{sec:models}

Our goal is to elucidate the connections between the optically thin \oi~135.6~nm line and the core of the \mgii~k line, especially as concerns the spectral widths of these lines and their connection to the chromospheric plasma, in particular the velocity field and related quantities such as the magnetic field, density, ionization state, and radiation field. In order to do so we will first show that the formation of these two lines occurs in roughly the same chromospheric height range and thereafter show how the line parameters respond to different simulated solar scenes.  

\begin{figure*}
    \centering
    \includegraphics[width=0.9\linewidth]{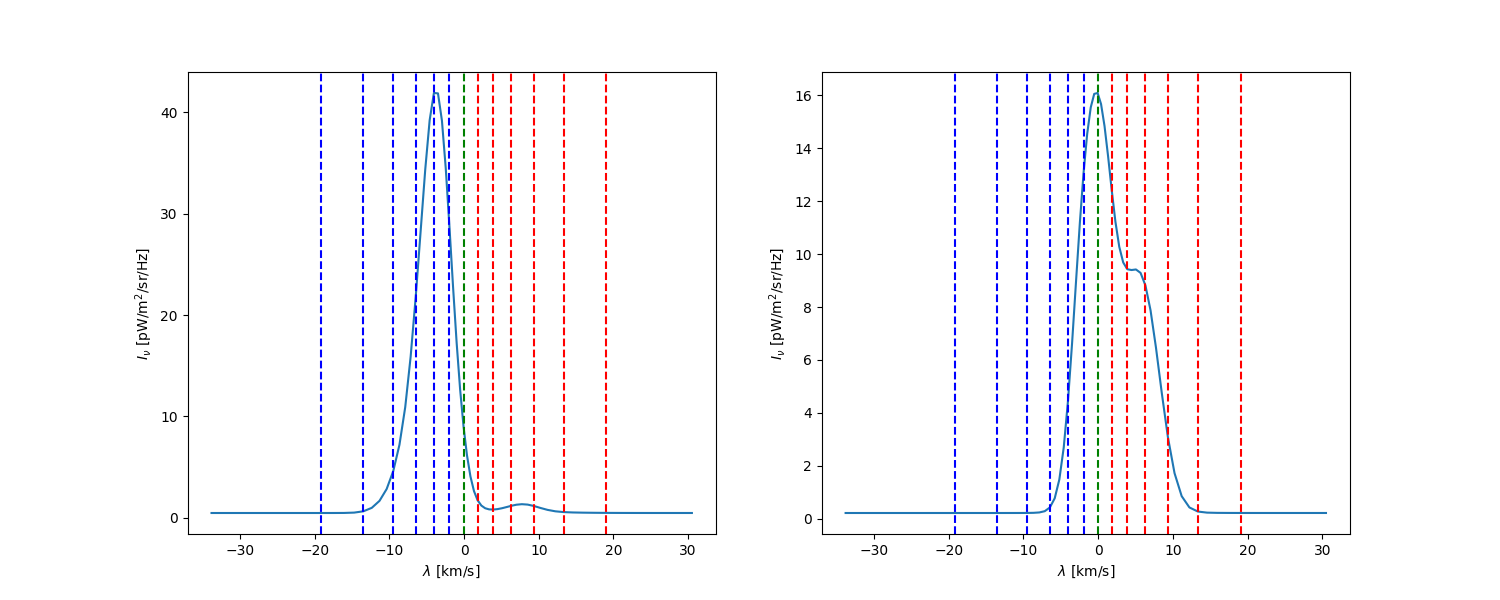}
    \caption{\oi~135.6~nm spectra (full blue lines) for two random locations in the plage model. The dashed green line shows the central wavelength $\lambda_0$ while the dashed blue and red vertical lines show the wavelengths at which the opacity $\chi_\nu(z)$ and source function $S_\nu(z)$ have been saved.}
    \label{fig:oi_chosen_wvl}
\end{figure*}

\begin{figure*}[h!]
    \centering    \includegraphics[width=0.9\linewidth]{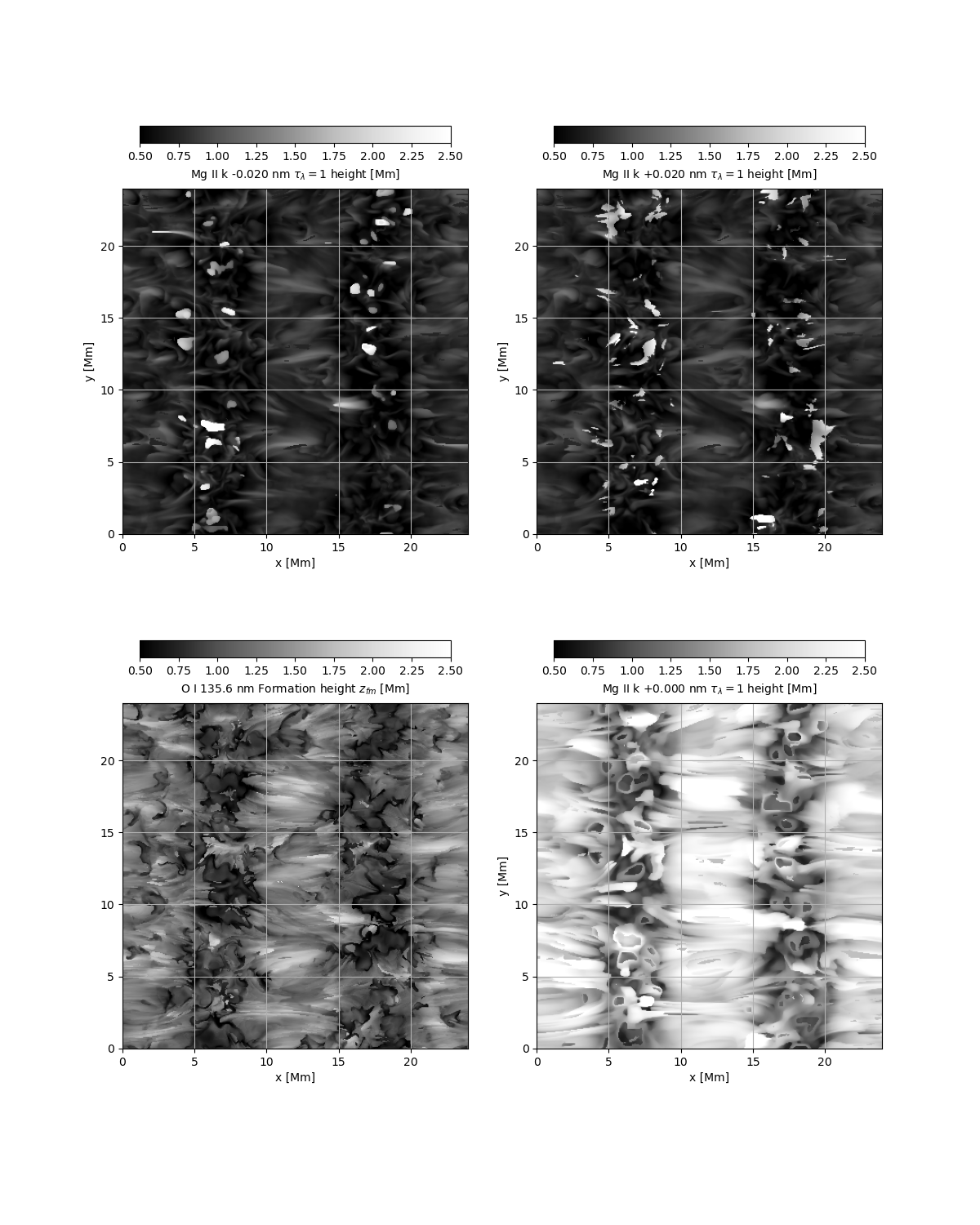}
    \caption{Height of formation for \oi~135.6~nm (lower left) and three wavelengths in the core of the \mgii~k~279.635~nm line; at $\delta\lambda = [-0.02, 0, 0.02]$ nm (upper left, lower right, upper right respectively) in the non-equilibrium hydrogen ionization plage model. Note the strong similarity of the structures in the oxygen and ``wings'' of the \mgii~core close to the $k_2$ peaks, The $k_3$ peak structures are dissimilar and is clearly formed at greater heights than the \oi~line.  The \mgii~wings are formed at lower heights than \oi~135.6~nm. Note also the regions where the wings are formed at heights $> 2$~Mm; these are sites of large upward or downward flows (of order $> \pm 20$ km/s) that shift the higher opacity $k_3$ peak to these wavelengths. }
    \label{fig:mgii_oi_heights}
\end{figure*}

\begin{figure*}[h!]
    \centering
    \includegraphics[width=0.9\linewidth]{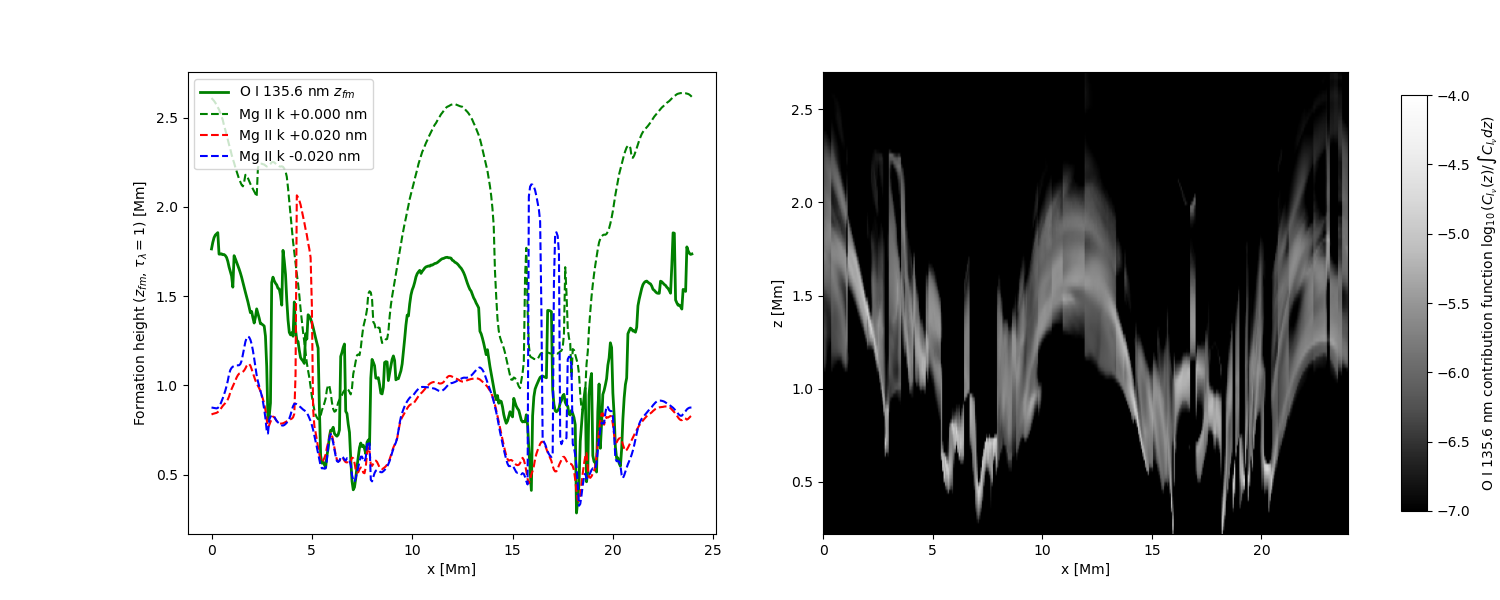}
    \caption{Left panel: Formation height $z_{\rm fm}$ of the \oi~ 135.6~nm line (solid green) at a random position $y = 7.0$~Mm as a function of $x$. Overplotted are the formation heights, $\tau_\lambda=1.0$, of the $k_3$ peak (dashed green), and close to the $k_{2r}$ (dashed red) and $k_{2v}$ (dashed blue) peaks: The \oi~line is formed in between the formation heights of these \mgii $k$ profile locations. 
    Right panel: Contribution function to the \oi~135.6~nm line at the same position $y = 7.0$ Mm as function of $x$.}
    \label{fig:mgii_tau_oi_contribution}
\end{figure*}

\subsection{Numerical model: Bifrost code\label{sec:br_code}}
In the following we present a number of models constructed with the Bifrost code \citep{2011A&A...531A.154G} which solves the radiative MHD equations in a computational box spanning the upper convection zone to the corona. Optically thick radiation is treated according to the methods presented by \citet{2000ApJ...536..465S} which allows the treatment of scattering in the lower chromosphere using four opacity bins. Radiative losses in the upper chromosphere, transition region and corona are treated according to the methods described by \citet{2012A&A...539A..39C}, while thermal conduction along the magnetic field is included.

All these models are constructed by setting the entropy inflow at the bottom boundary to a level that ensures the effective temperature is roughly equal to the solar value of $T_{\rm eff} = 5780$~K and with an equation of state that contains the most important electron donors found on the Sun at solar abundances. Magnetic fields can be introduced at the bottom boundary in the shape of sheets or slabs and will enter the computational domain in regions of inflow. The upper boundary is set to be transparent by formulating the MHD equations in characteristic form and setting the incoming characteristic quantities (waves) to zero.

The models differ in the depths they extend to in the convection zone and in the coronal heights they reach. Perhaps more significantly, the models differ in the following ways: spatial resolution, whether the equations of statistical equilibrium are solved for hydrogen, or whether LTE is assumed in calculating the hydrogen ionization and excitation state, and finally the initial and imposed magnetic field topology and strength. 

\subsection{Formation of the \oi~135.6~nm line \label{sec:oi_formation}}

In the following we will compute \oi\ and \mgii\ intensities mainly using the 1.5D RH code \citep{2001ApJ...557..389U,2015A&A...574A...3P}, but for \mgii\ synthesis we will also occasionally use the fully 3D Multi3D code \citep{Leenaarts_Carlsson:2009}. In order to synthesize these lines we need both numerical and atomic models; the former models are described below, while for the latter we use the \oi~atomic model developed by \citet{2015ApJ...813...34L}, and the \mgii~atomic model described by \citet{2013ApJ...772...89L}.

Let us establish the height of formation ($z_{\rm fm}$) for \oi~135.6~nm and its relation to the \mgii~k line by considering synthetic spectra taken from the plage model. This model, described in section \ref{sec:plage}, is one where hydrogen ionization is calculated via the time dependent rate equations --- henceforth called non-equilibrium hydrogen ionization (NEQ). The height $z_{\rm fm}$ varies from model to model, depending on magnetic topology and strength.

We use the contribution function to intensity 
\begin{equation}
C_{I_\nu}(z) = S_{\nu}\exp(-\tau_\nu)\chi_\nu
\end{equation}
\noindent where $S_\nu$ is the source function, $\tau_\nu$ the optical depth and $\chi_\nu$ the opacity at frequency $\nu$. The height of formation is then given by 
\begin{equation}
z_{\rm fm} = \int z\, C_{I_\nu(z)}dz
\end{equation}
\noindent where the integral is calculated at the frequency, $\nu$, of maximum emission $I_\nu$. In order to carry out this integration, we save the source function, the optical depth, and the opacity as functions of height at 13 spectral positions centered on 135.6~nm. As shown in Figure~\ref{fig:oi_chosen_wvl}, these are used to calculate and characterize the \oi~spectral profile. 

The resulting formation heights $z_{\rm fm}$ and contribution function $C_{I_\nu}$ are shown in Figure~\ref{fig:mgii_oi_heights} and Figure~\ref{fig:mgii_tau_oi_contribution}. The \mgii~k line heights are characterized by the height at which $\tau_\lambda = 1$ for three wavelengths: at the line core $\lambda_0 =  279.6352$~nm (k$_3$) and at $\lambda_0\pm 0.02$~nm. The latter two wavelengths roughly coincide with the average wavelengths of the k$_{2r}$ and k$_{2v}$ peaks. The maps of these heights have very similar morphologies, as seen in the upper two panels of Figure~\ref{fig:mgii_oi_heights}. The lower left panel of Figure~\ref{fig:mgii_oi_heights} shows that the formation height of the \oi\ line varies between 0.5 and 1.75~Mm above the photosphere, which lies between the heights found for the \mgii\ k$_2$ peaks and k$_3$. Morphologically, the map of the \oi\ heights corresponds more with the maps found for the k$_2$ peaks, and we will later see that this is also true for the intensities. In general, for all models, we find that the \oi\ 135.6~nm line reflects \mgii\ height formation at roughly $\lambda_0\pm 0.015$~nm. The k$_2$ heights in the upper two panels show some select locations that are formed much higher, with $\tau_\lambda = 1$ heights $> 2.5$~Mm. These are all locations where velocities are large enough, $u_z > \pm 20$~\kms, that the higher opacity core of the line is Doppler shifted beyond $\pm 0.02$~nm. (We have plotted the \mgii\ heights at $\pm0.02$~nm instead of $\pm0.015$~nm to decrease the number of such locations and thereby make the images clearer, the latter positions are formed a few hundred kilometers above the $\pm0.02$~nm locations).

That the formation of the \oi~135.6~nm line occurs between the heights where the \mgii~k$_2$ peaks and the k$_3$ core occur is confirmed by inspection of Figure~\ref{fig:mgii_tau_oi_contribution} which shows the formation heights at a specific $y$-position as a function of location along the $x$-axis. Again, we find that the heights of the k$_2$ peaks are very similar (except in locations where the velocities and hence Doppler shifts are larger than $\pm 0.02$~nm) at $z\simeq 1$~Mm. The k$_3$ core is, at least for this model, formed significantly higher, sometimes surpassing $z\approx 2.0$~Mm. The contribution function shown in the right panel shows that the \oi~line is formed between these heights, hence the non-thermal ($1/e$) width of this line should give an accurate representation of the non-thermal velocity structure forming the core of the \mgii~line as discussed by \citet{2023ApJ...959...87C}.
\section{Results \label{sec:results}} 

\subsection{Quiet Sun model}\label{sec:qs}

A number of quiet Sun models have been calculated. These models all extend from 8~Mm below the photosphere to 52~Mm above the photosphere vertically and cover $72\times72$~Mm$^2$ horizontally. In the vertical $z$-dimension the grid size is 20~km in the photosphere, chromosphere, transition region and corona, but increases to roughly 100~km at greater depths in the convection zone and at greater heights in the corona, since pressure scale heights are larger there. The horizontal resolution is 100~km, except for in the {\tt qs072050} and {qs072050\_x2s} models where it has been increased to 50~km. 

\begin{figure*}
    \centering  
\centerline{\includegraphics[width=1.2\linewidth]{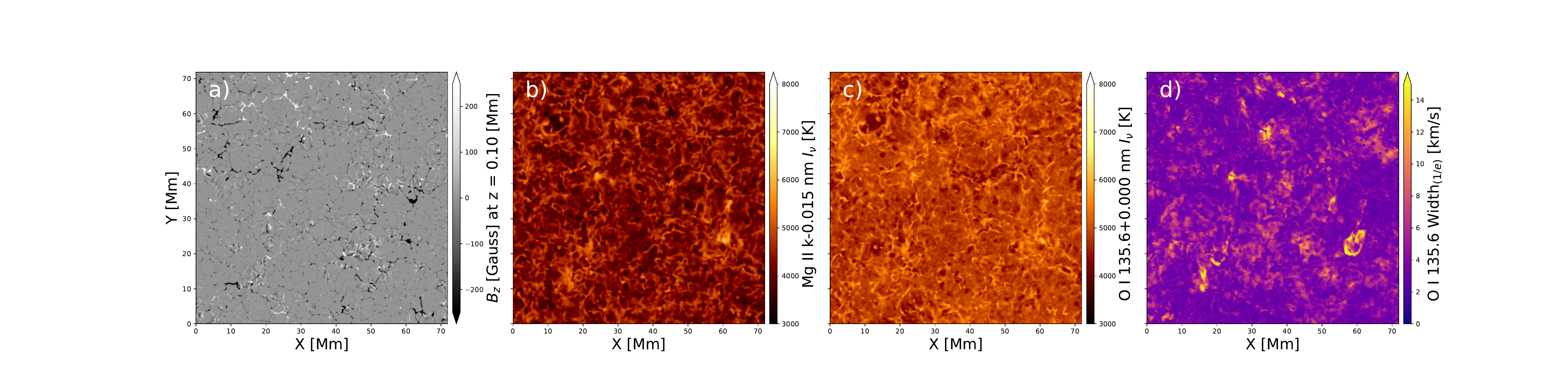}}
\vspace{-0.6cm}
\centerline{\includegraphics[width=1.2\linewidth]{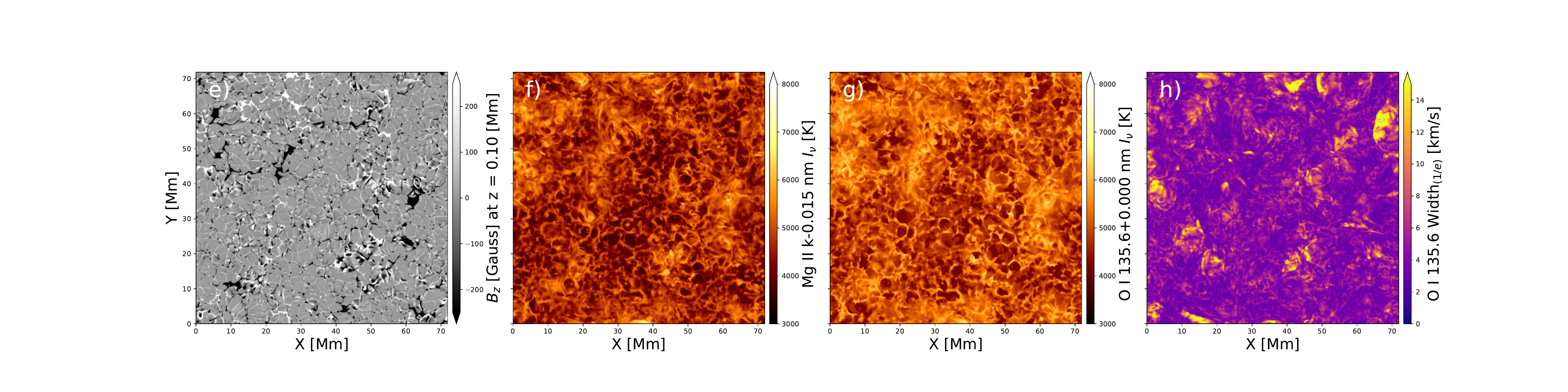}}
    \caption{Quiet Sun models. The upper row shows the weakest field model, while the lower row shows the stronger field model; a) and e): vertical magnetic field at $z=0.1$~Mm, saturated to [-250,250]~Gauss.  b) and f): \mgii~k line radiation temperature at $-0.015$~nm from line center, close to the k$_{2v}$ peak.   c) and g): Line center intensity of the \oi~135.6~nm line. d) and h): \oi~135.6~nm $(1/e)$ widths. Note the close resemblance of the \mgii~and \oi\  intensities. }
    \label{fig:qs_model}
\end{figure*}

Let us begin with the weakest field quiet Sun model {\tt qs072100\_d2n}. In the photosphere, at $z=0.1$~Mm, roughly the height of \fei~637.1~nm formation, this model has an average unsigned vertical field strength of $\langle|B_z|\rangle=17.13$~Gauss, and a mean vertical field of $\langle B_z\rangle = 6.96$~Gauss. Vertical photospheric field strengths vary from $-1640$~Gauss to $1597$~Gauss in the strongest patches. The field is distributed as seen in the upper left panel, a), of Figure~\ref{fig:qs_model}. These numbers are very close to what \citet{2025ApJ...990..195G} report from Swedish 1~metre Solar Telescope observations for the quiet Sun internetwork areas, but are considerably lower than what \citet{2014ApJ...789..132R,2016A&A...593A..93D} predict to be consistent with high resolution modeling and observations; namely $\langle |B_z|\rangle\simeq 60$~Gauss. We construct models with field strengths lying between these values.

\setcounter{footnote}{0}
Panel b) of Figure~\ref{fig:qs_model} shows a spectrogram  of the \mgii~k line core $-0.015$~nm to the blue of line center; to the right, in panel c), the central line intensity of the \oi~135.6~nm line is displayed. (We use the radiation temperature to plot intensities in this paper $T_{\rm rad}= ({h\nu/k_B})/\log_{10}\left[{2h\nu^3/(c^2I_\nu)+1}\right]$ for better display purposes, as the intensity varies very rapidly with varying radiation temperatures at UV and EUV wavelengths.) Finally, the right panel, d), shows the $(1/e)$ half width of the \oi~135.6~nm line as calculated from the second moment of the intensity:
\begin{equation}
\sigma_{\rm model} = \sqrt{\frac{\int I_\lambda (\lambda - \lambda_0)^2d\lambda}{\int I_\lambda d\lambda}-u_{1st}^2}
\end{equation}
where the integral is calculated over a wavelength span of $d\lambda = [-0.05,0.05]$~nm and we have subtracted a linear approximation to the continuum, using the intensity at the boundary points of this range, while $u_{1st}$ is the first moment of the intensity. The $(1/e)$ half width is then related to the second moment of the intensity via $w_{(1/e)} = \sqrt{2}\,\sigma_{\rm model}$.

Panels e) - h) show the same parameters for the {\tt qs072100\_x2s} model which has a significantly stronger photospheric field: At $z=0.1$~Mm a field strength of $\langle|B_z|\rangle=62.51$~Gauss and a mean vertical field of $\langle B_z\rangle = 2.56$~Gauss. The vertical photospheric field strengths vary from $-1505$~Gauss to $1758$~Gauss in the strongest patches.

 Though differing in radiation temperature, we find that the \oi\ line has a very similar morphology to that of the \mgii\ line close ($\delta\lambda = \pm 0.015$~nm) to line center. This is most clearly seen in the stronger field model, which also shows that both intensities are strongest in locations of strong magnetic field. The similarity was also found in IRIS observations and noted by \citet{2023ApJ...959...87C}. 

We note that the pattern of oxygen line widths shown in the left panels d), and h) display variation across the entire field of view; large widths often, but not always, highlighting regions in the vicinity of strong photospheric flux concentrations. Regions with smaller line broadening, which cover most of the field of view, show amorphous 1--5~Mm sized regions spread seemingly at random. There is no obvious correlation between the \oi~line intensity and width, but see below.

\subsubsection{\oi~135.6~nm line statistics}
On average we find the \oi~line intensities in the weak field {\tt qs072100\_d2n} model to be lower than what is found in typical IRIS quiet Sun \oi~profiles  --- line peaks of 0.13~\pW\  vs 0.25~\pW\ \citep[see][]{2015ApJ...813...34L}. The same is true for the \mgii~k line intensities which are weaker, by a factor 2 or so, than what IRIS quiet Sun observations show. (This is shown in Figure~\ref{fig:qs_mg_width} for a very large dense raster obtained by IRIS on February 25, 2014 at 18:59:47 UTC.)

\begin{figure*}
    \centering
    \includegraphics[width=0.9\linewidth]{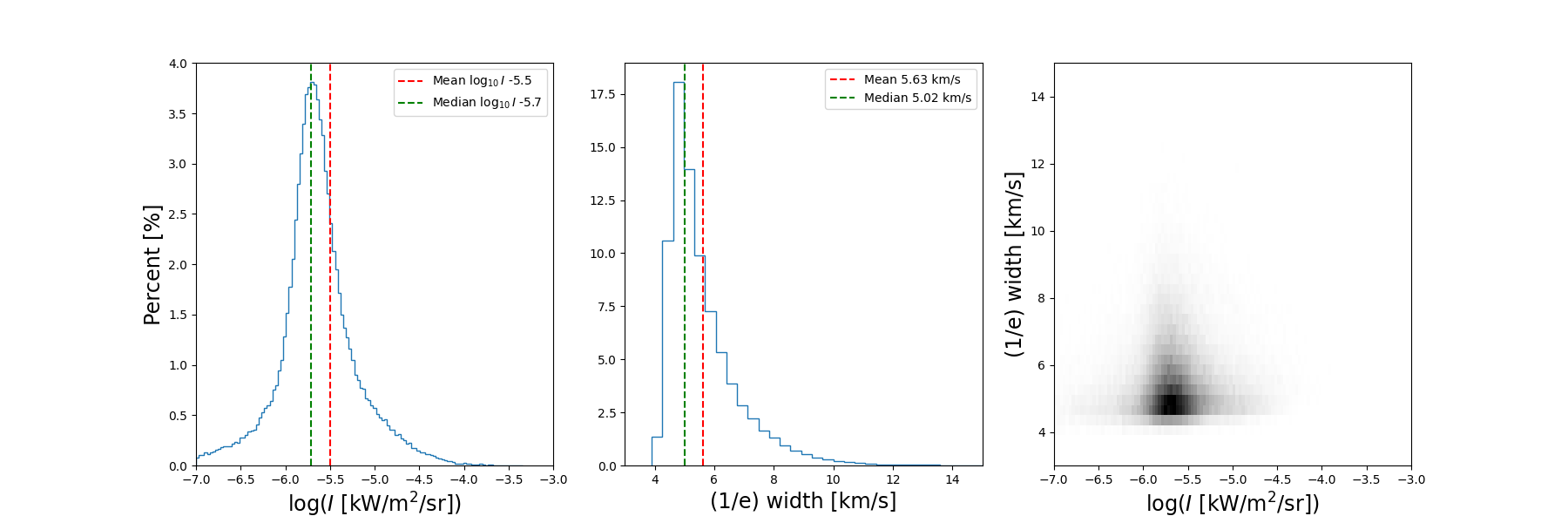}
    \includegraphics[width=0.9\linewidth]{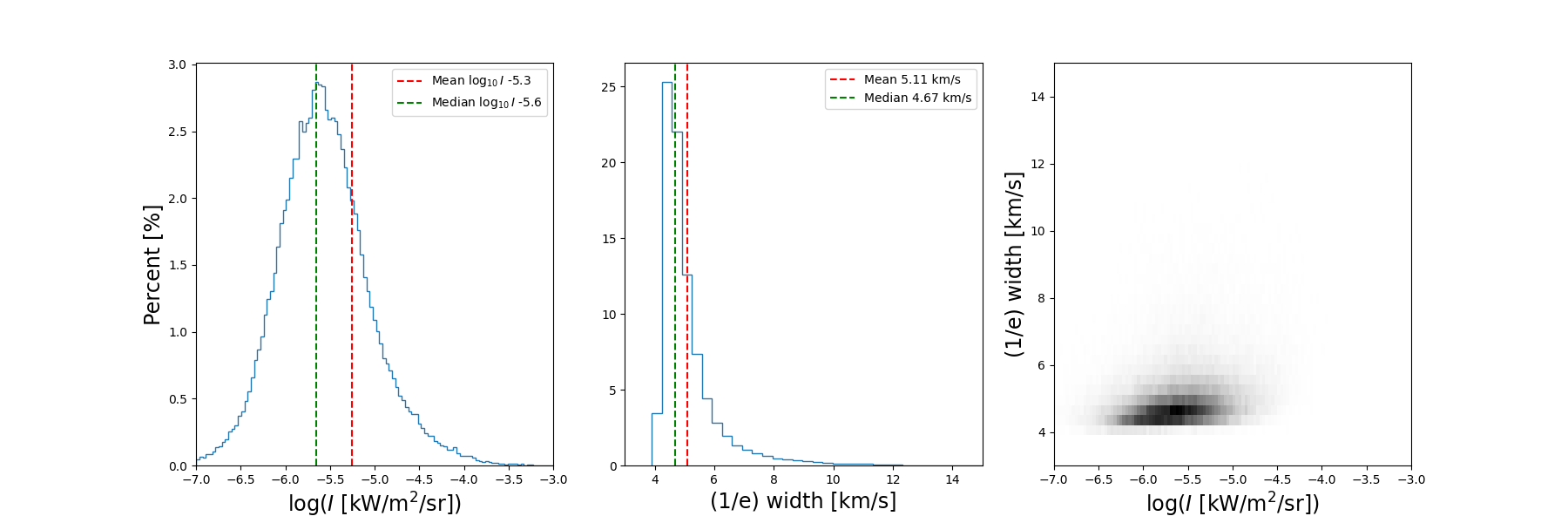}
    \includegraphics[width=0.9\linewidth]{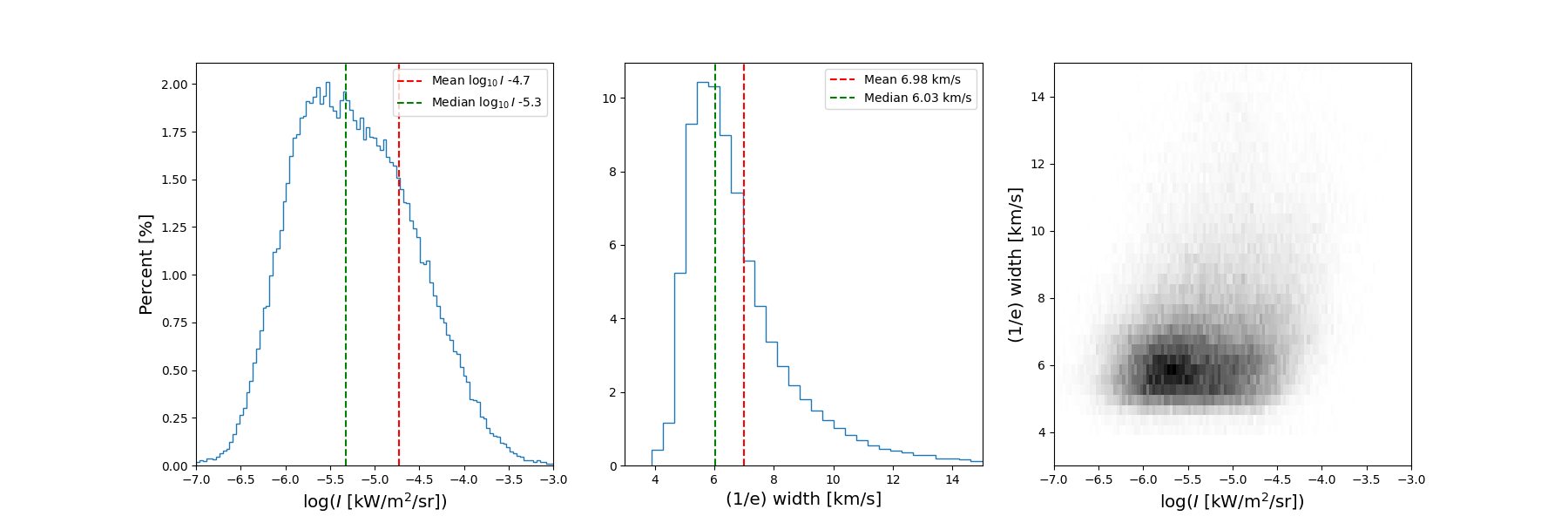}
    \caption{Histograms of the total line \oi~intensities (left panel), $(1/e)$ line widths (central panel), and the JPDF of these quantities. The top row shows the weakest field quiet Sun model, the middle row the medium field strength model, and the bottom row the strong field model. These models differ in their average photospheric vertical magnetic fields $\langle|B_z|\rangle$ which are 17.13~ Gauss, 27.69~Gauss, and 65.13~Gauss respectively. The red dashed lines in the left and central panels shows the mean intensities and widths respectively, while the green dashed lines shows the median values.}
    \label{fig:oi_histo_jpdf}
\end{figure*}

The statistics of total \oi~line intensities and widths are shown in Figure~\ref{fig:oi_histo_jpdf}.  The first moment of the intensity, for this weak field model, ($\int I_\nu d\nu$), ranges from $10^{-7}$ to $10^{-4}$~kW/m$^2$/sr, with a median of $\log_{10} I = -5.7$ and a mean of $-5.5$, as shown in the upper left panel.  The upper central panel shows that  the synthesized $(1/e)$ widths in this model range from 4~\kms\ up to $\sim 15$~\kms\ with a median width of 5.0~\kms\ and an average of 5.6~\kms. In this figure, and in the following similar figures, we have added the IRIS instrumental $(1/e)$ broadening of 3.4~\kms, which adds of order 1~\kms to the total width:
$\sigma = \sqrt{\sigma_{\rm inst}^2+\sigma_{\rm model}^2}$. In addition we have spatially convolved the data with a gaussian with FWHM of $0.33$~arcsec, and, since the \oi~widths measured used an observational program that sums over 2 spatial pixels of size 0.16\arcsec, with macropixels of $0.33$~arcsec. This is done in order to facilitate comparisons with IRIS data described by \citet{2023ApJ...959...87C}. The spatial summing has only a modest effect, adding less than $0.5$~\kms\ to the line widths.

The upper right panel shows the JPDF of the total line intensity and $(1/e)$ width: for this weak field model there is no obvious correlation between these quantities.

\begin{figure*}   
    \centering
    \includegraphics[width=0.8\linewidth]{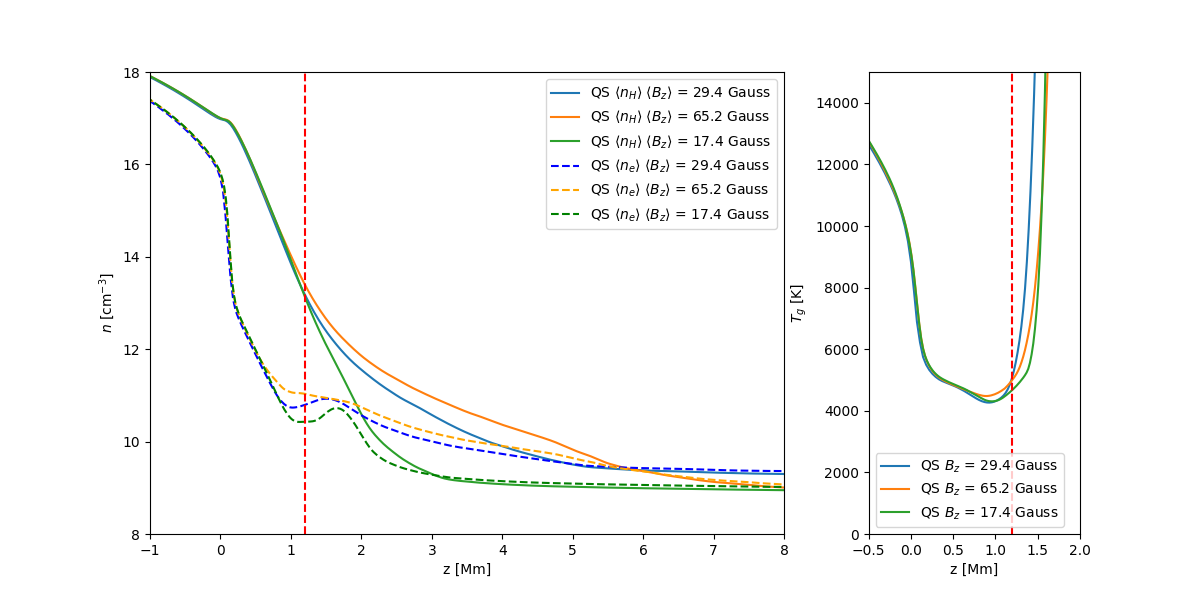}
    \caption{Mean electron ($n_{\rm e}$) and total hydrogen ($n_{\rm H}$) densities as a function of height $z$. The red dashed lines shows the approximate height of the formation of the \oi~line $z_{\rm fm}$.}
    \label{fig:QS_n_Tg}
\end{figure*}

The lower two rows of Figure~\ref{fig:oi_histo_jpdf} show how these numbers change as we double the average magnetic field strength to $\langle|B_z|\rangle\simeq 28$~Gauss and then double again to $\langle|B_z|\rangle\simeq 62$~Gauss. Consider first the intensities. The first doubling changes neither the median, nor the mean, to any significant degree; indicating that, on average, the magnetic field does not play a major role in chromospheric structure at the heights of formation of the \oi~line for the two weakest field models. On the other hand, when doubling again to the model with $>60$~Gauss photospheric fields, we find that both the median and the mean intensities have increased by a factor of roughly 4. Since the \oi~line intensities are tightly coupled to the electron density $n_{\rm e}$ where the line is formed, we expect to find higher average densities in the strongest field case. Indeed, computing the electron density at the formation height of the \oi~line, $z_{\rm fm}$, as 
\begin{equation}
n_{\rm e,fm} = \int n_{\rm e} C_{I_\nu(z)}dz
\end{equation}
\noindent
we find in the weaker field models average electron densities of $1.7\times 10^{17}$~m$^{-3}$ and $1.6\times 10^{17}$~m$^{-3}$ respectively, while in the strongest field model we find $\langle n_{\rm e,fm}\rangle = 3.0\times 10^{17}$~m$^{-3}$, almost two times higher than in the weaker field models. This significant increase in \oi\ intensity and its dependence on electron density is as expected since the \oi~line intensity is predicted to be closely proportional to $n_{\rm e}^2$ as shown by \citet{2015ApJ...813...34L}. Further evidence of this difference can be found when considering the horizontal mean of the electron and total hydrogen densities as a function of height, as shown in Figure~\ref{fig:QS_n_Tg}: In the strong field model both densities remain higher than what is found in the weaker field models from $z=0.5$~Mm above the photosphere and remain higher up to $z \simeq 5$~Mm. 

\begin{figure}   
    \centering
    \includegraphics[width=0.9\linewidth]{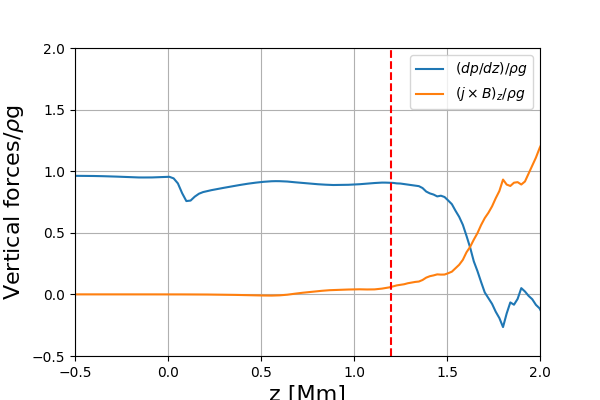} \\
    \includegraphics[width=0.9\linewidth]{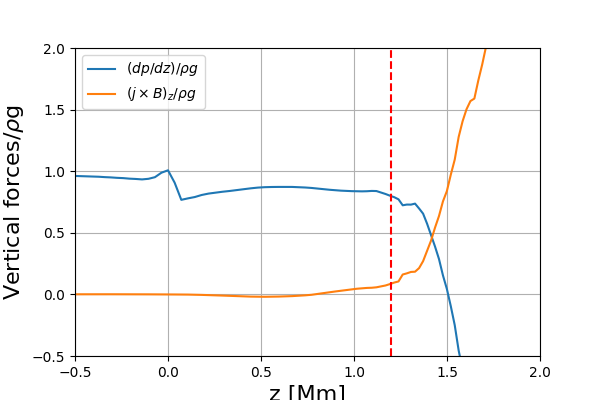}\\
    \includegraphics[width=0.9\linewidth]{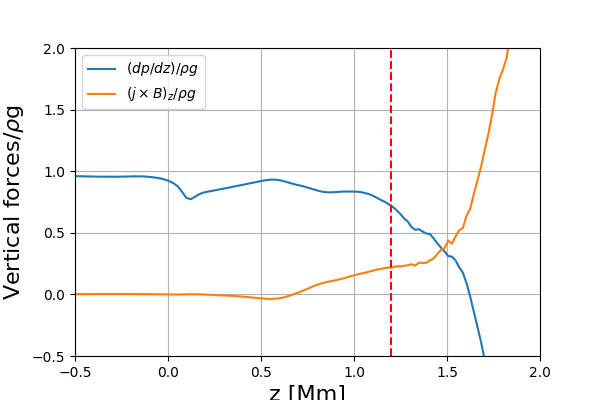}
    \caption{Mean force of vertical pressure gradient ($dp/dz$) and Lorentz force $(\mathbf{j\times B})_z$, both normalized to the gravitational force, as functions of height for the weak ($\langle|B_z|\rangle =17.3$~Gauss, top panel), medium ($\langle|B_z|\rangle=27.69$~Gauss) and strong ($\langle|B_z|\rangle=65.2$ Gauss, lower panel) magnetic field model as functions of height $z$.
    As in figure~\ref{fig:QS_n_Tg} the red dashed lines show the approximate height of the formation of the \oi~line $z_{ \rm fm}$.} 
    \label{fig:QS_Fpz_FBz}
\end{figure}

This difference could be due to the slightly higher average temperatures found above $z=0.5$~Mm for the strong field model, which implies a larger pressure scale height. However, as seen in Figure~\ref{fig:QS_Fpz_FBz}, we also find that the vertical component of the Lorentz force $(\mathbf{j\times B})_z$ rises rapidly above $z=0.5$~Mm, becoming equal to the vertical pressure gradient force at $z=1.5$~Mm. 
In the weaker field models the transition to an average force balance where the magnetic component plays a small but measurable role comes at a greater height, at $z \simeq 1$~Mm, closer to or in the transition region. Above $z\simeq1.5$~Mm the magnetic component of the force balance dominates gravity in all three models.

\begin{figure*}   
    \centering
    \includegraphics[width=0.9\linewidth]{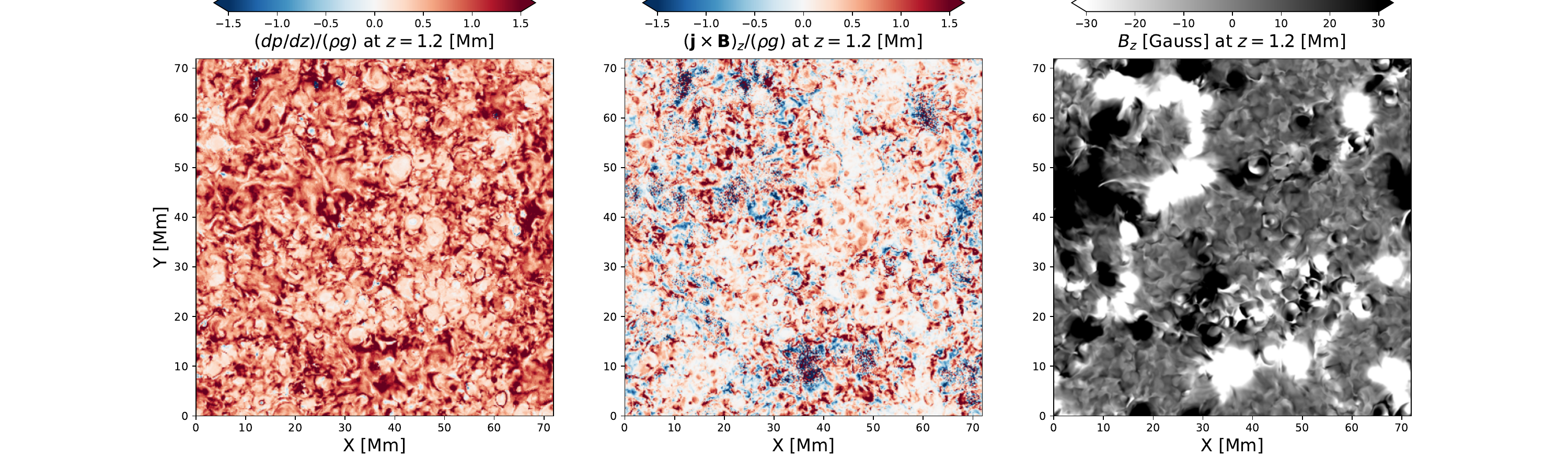} 
    \caption{Mean force of vertical pressure gradient ($dp/dz$) and Lorentz force $(\mathbf{j\times B})_z$, both normalized to the gravitational force, at the average height of \oi~line formation ($z=1.2$~Mm for the strong ($\langle|B_z|\rangle=65.2$~Gauss) Quiet Sun  model.
    The right panel shows the vertical component of the magnetic field at the same height.}
    \label{fig:QS_Fpz_FBz_Bz}
\end{figure*}

\begin{figure*}   
    \centering  \includegraphics[width=0.9\linewidth]{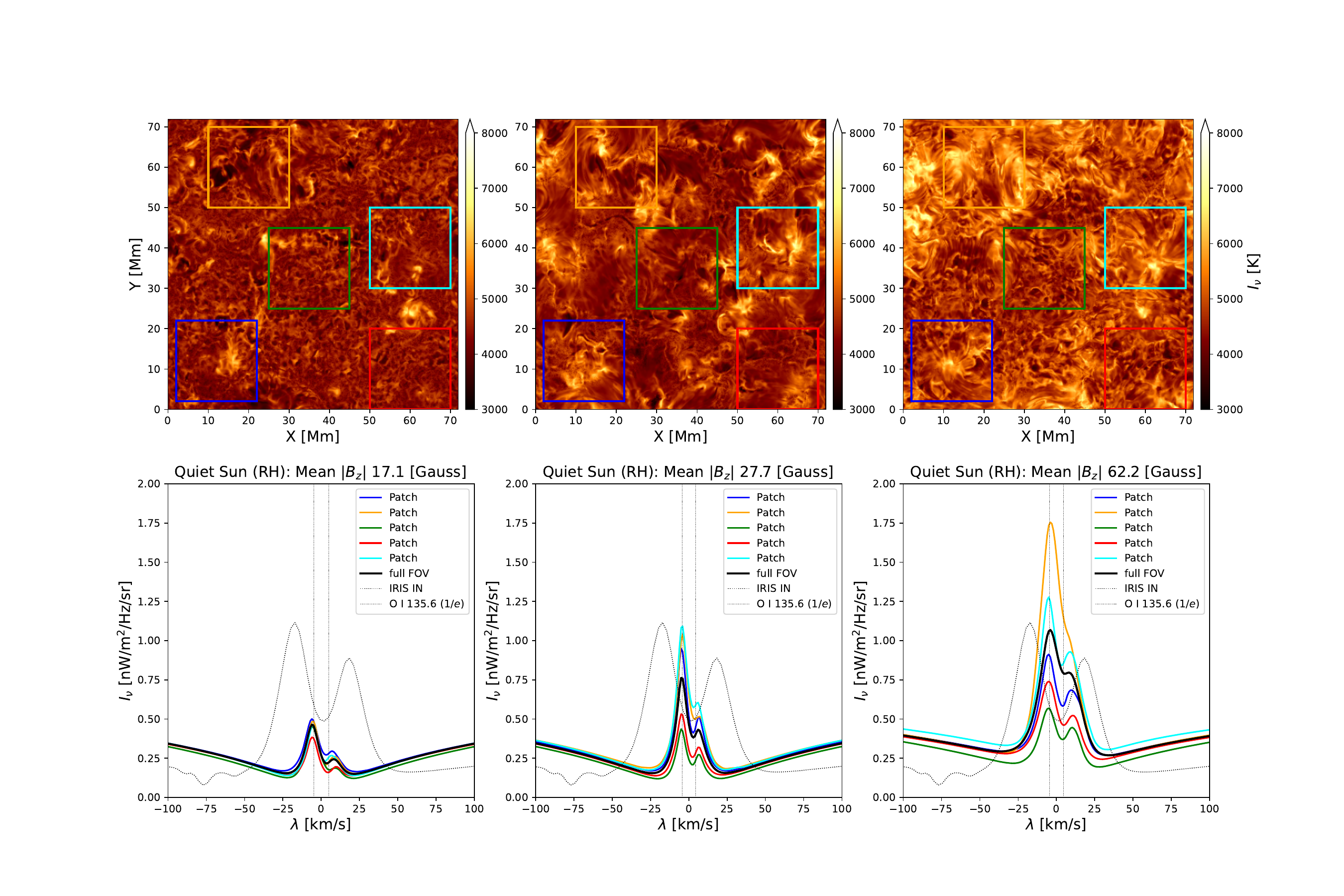}
    \caption{\mgii~core line widths for the {\tt qs072100\_d2n}, {\tt qs072100}, and {\tt qs072100\_x2n} models. The average profiles are shown for five, more or less randomly selected patches, in color, while the average profile for the entire FOV is shown in black. Overplotted in dashed black is an average IRIS-observed internework profile taken with a very large dense raster on February 25, 2014 at 18:59:47 (UTC). The $(1/e)$ widths of the corresponding \oi~line profiles are indicated by the dotted vertical lines.}
    \label{fig:qs_mg_width}
\end{figure*}

Further details of this force balance may be gleaned on inspection of Figure~\ref{fig:QS_Fpz_FBz_Bz} which shows the pressure gradient and Lorentz forces at $z_{\rm fm}=1.2$~Mm for the strongest field quiet Sun model. At this height the average pressure gradient pushes upwards with 75\%,  while the average Lorentz force accounts for 25\% of the resistance to gravitational acceleration. These numbers are far from constant at smaller scales. The pressure gradient is essentially always directed upwards at this height, with only a few small downward directed points. On the other hand, the Lorentz force shows both upwards and downward directed regions, strongest in regions of heightened magnetic fields. Viewing the figure, it is clear that these forces do not consistently sum to  counteract gravity, implying that the chromosphere is pervaded by dynamics (waves) with amplitudes of order the speed of sound.  

% zfm = 1.25 Mm for x2 model, 1.16 Mm for d2 model, and 1.25 Mm for "standard" model.

Thus, the density structure of the three quiet Sun models presented show significant differences as the average magnetic field strength is varied. This impacts the total \oi~line intensity, and as we shall see below, also the \mgii~k line intensities. However, such differences are not as clearly found in the widths of the \oi~line, where figure~\ref{fig:oi_histo_jpdf} shows median values of $5.0$, $4.67$ and $5.54$~\kms, and mean values of $5.63$, $5.11$, and $6.49$~\kms\  when ranging from the weakest to the strongest field models. We do see that there is a slight tendency for the widths to increase in the strongest field model, and also that this model's JPDFs show a slight correlation between \oi~line intensity and line width. 

This finding is consistent with observational findings of \citet{2023ApJ...959...87C} who found extremely low widths in inter-network regions: In several locations, so
low as to be compatible with a lack of non-thermal
broadening.  They also find significantly larger line widths in the vicinity of the brighter network regions, where magnetic fields are stronger.

To summarize, we find that increasing the average vertical magnetic field $\langle|B_z|\rangle$ in the photosphere from 17~Gauss to 62~Gauss increases the scale height and the electron density in the chromosphere, with consequences for the total intensity of the \oi~135.6~nm line, but find little change in the line width statistics and therefore presumably few changes in chromospheric velocities near $z=1$~Mm, except near stronger field network regions where widths and presumably turbulent velocities are greater.

\subsubsection{Implications for the \mgii~line core}
Consider now how the changes in magnetic field strength impact the \mgii~k line, and in particular the line core width. In the following, when setting a numerical value to the width of the core, we fit a gaussian to the body of the line core, then subtracting an ``interior'' gaussian to model the wavelengths surrounding the k$_3$ peak (unless the profile is single peaked). We then define the width of the \mgii\ core as the $(1/e)$ width of the ``outer'' gaussian. Figure~\ref{fig:qs_mg_width} shows the relationship between magnetic field strengths and \mgii~line profiles for the models discussed above. The top row shows how the spatial structuring of the map of \mgii~k$_3$ line core radiation temperature changes and becomes more complex as the average $\langle |B_z|\rangle$ magnetic field increases from the leftmost to the rightmost panel. We have chosen 5 patches, more or less at random locations, to illustrate these changes. All three models share a similar relationship with the magnetic field, showing enhanced emission and 10-30 Mm long filamentary shapes where the magnetic fields are strong. However, the strongest field model has many more such regions. Weaker field regions in all three models show structures reminiscent of reverse granulation, but on a larger scale than granules appear in the photosphere. 

The bottom row shows the impact of this increase on the \mgii~line core profiles. A typical average IRIS quiet Sun internetwork line profile is plotted for comparison. First of all, we note that the effective temperature of this model is higher than the solar value, and hence that outside the line core the intensities are higher in the models than on the Sun. However, in the core of the weakest field model, core intensities are at least a factor 2 smaller than what is observed. Furthermore, the core width is significantly narrower than what the observations show, some 8.3 vs typically 20~\kms\ in observations. These results are consistent with what has been found and reported in  earlier studies \citep[e.g.][]{2023ApJ...944..131H}.  When the average magnetic field increases, core intensities increase correspondingly. The 27~Gauss field model intensities found in the central panels of Figure~\ref{fig:qs_mg_width} show that in all chosen patches, as well as for the average field of view, the intensities increase. In the cyan patch the intensity increases to a maximum k$_{2v}$ intensity of 1~\nW, close to typical IRIS inter-network values, though the average intensity over the full field of view is still well below observed average intensities. On the other hand, the intensities in the strongest 62~Gauss field model show core intensities very similar to what is observed, between 0.5 and 1.5~\nW\ for quiet Sun. 

Perhaps more striking is the evolution of the core width with increasing average photospheric magnetic field strengths. In the weaker field models the \mgii\ core remains narrow, even as the average field strength increases from 17 to 27~Gauss, rising only from 8.3 to 9.5~\kms. However, a dramatic broadening of the profile is found when the average field strength reaches $>60$~Gauss, and while still being too narrow, is measured to be 14.5~\kms, much closer to the observed core width than the other models shown.

Note that this does not imply that non-thermal velocities are unimportant in setting the \mgii~core width nor in setting the chromospheric structure and dynamics. Indeed, as implied by \citet{2024A&A...692A...6O} the turbulent pressure ($\rho w_{\rm turb}^2$) is of the same order of magnitude, as the pressure ($\sim\rho c_s^2$) is at the height of \oi~line emission ($z_{\rm fm}$). 

This should not be surprising, as both the observed and modeled \oi~widths are only slightly smaller than the local speed of sound. Subtracting the contributions from the instrumental and thermal widths, as well as the effects of spatial smearing, which summed together are of order 2.5~\kms\, we arrive at an approximate non-thermal velocity of 4.5~\kms\ for the quiet Sun models. Using this relatively conservative measure gives an average turbulent pressure of 1.6~\Nmtwo, while the average thermal pressure is only slightly larger: 2.3~\Nmtwo\ at the average height of \oi~line formation $z_{\rm fm}$. We therefore expect the turbulent pressure to play an important role in setting chromospheric structure. While the \oi\ line widths restrict how much Doppler broadening turbulent velocities can add in the \mgii\ k line core, these velocities can increase the chromospheric scale height and hence \mgii\ opacities. However, since the non-thermal widths do not increase dramatically with the average magnetic field strength --- at least for these quiet Sun models --- the turbulent pressure is clearly not the driving force that increases \mgii~line core widths as the magnetic field increases. 

A related question concerns the numerical resolution of the models presented. The 100~km horizontal grid size used here is fairly coarse, and while many aspects of e.g. photospheric dynamics are well represented at this resolution, others, such as granule and magnetic patch sizes, are not \citep[see][]{2025ApJ...990..195G}. One might therefore expect that increasing the spatial resolution will lead to smaller, perhaps more dynamic, phenomena leading to larger nonthermal chromospheric dynamics. On the other hand, the observed \oi\ line broadening places strong constraints on how much of the much larger \mgii~k line broadening can be due to velocities and Doppler broadening. 

We do find slightly larger \mgii\ widths in the 50~km horizontal grid size medium strength $\langle |B_z|\rangle\simeq 30$~Gauss ({\tt qs72050}) and strong $\langle |B_z|\rangle\simeq 60$~Gauss ({\tt qs072050x2s}) field models. However, the increase is limited: The measured $(1/e)$ width of the core rises from 9.5 to 10.7~\kms in the medium strength model, and from 14.6 to 15.2~\kms, when going from 100~km to 50~km resolution in the stronger field model. This effect is considerably smaller than what is found when increasing the magnetic field strength from $\sim 30$~Gauss to $\sim 60$~Gauss. Other examples of the effects of increasing the numerical resolution are shown in the following section~\ref{sec:cbp} in greater detail.

\subsection{Coronal Bright Point model}\label{sec:cbp}

\begin{figure*}
    \centering
\centering  
\centerline{\includegraphics[width=1.2\linewidth]{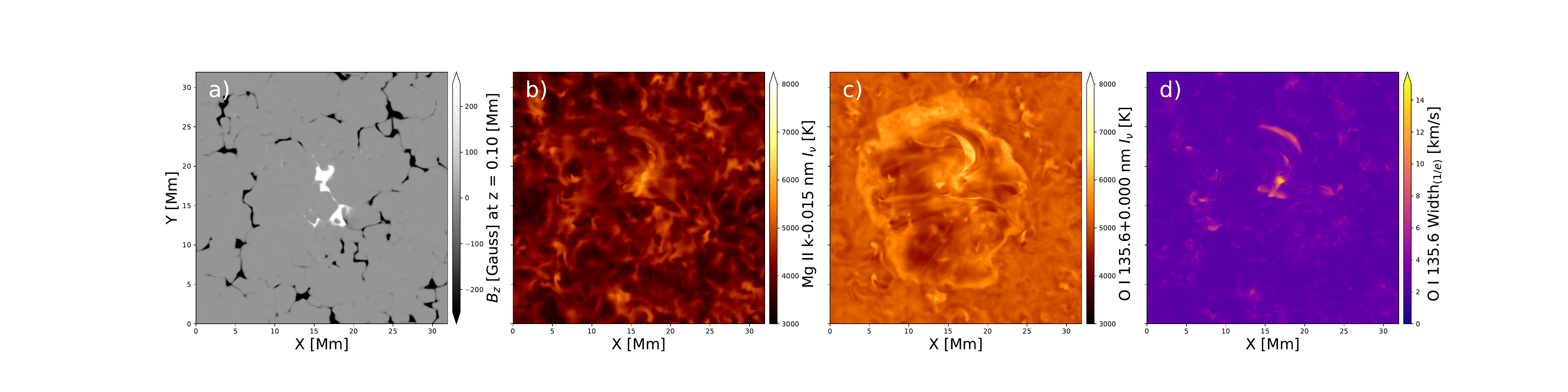}}
\vspace{-0.6cm}
\centerline{\includegraphics[width=1.2\linewidth]{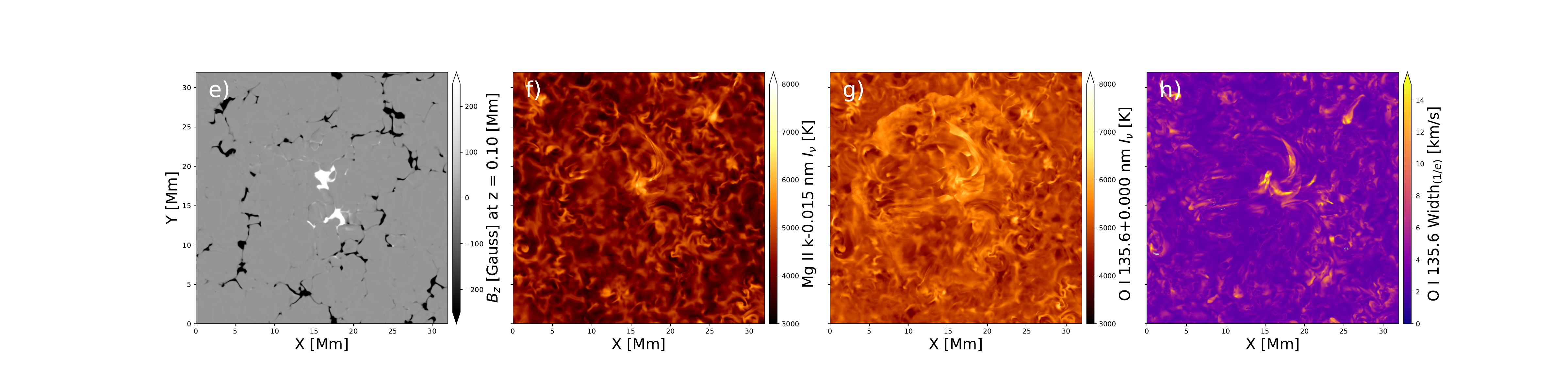}} 
    \caption{Coronal Bright Point (CBP) models. The upper row shows the low resolution CBP model, while the lower row shows the high resolution model; a) and e): vertical magnetic field at $z=0.1$~Mm, saturated to [-250,250]~Gauss.  b) and f): \mgii~k line intensity at $-0.015$~nm from line center, close to the k$_{2v}$ peak.   c) and g): Line center intensity of the \oi~135.6~nm line. d) and h): \oi~135.6~nm $(1/e)$ widths. }
    \label{fig:cbp_model}
\end{figure*}

%\dnstxt{
Two coronal bright point (CBP) models are presented here, both assume that hydrogen ionization is in LTE.
The first one, {\tt cbp032063}, models a CBP embedded within a coronal hole and it is described by \cite{2023ApJ...958L..38N}. 
The CBP topology consists in a fan-spine topology with a null point at 8~Mm in the corona.
The average unsigned vertical component $\langle|B_z|\rangle$ of the field is roughly 18~Gauss in the photosphere, while the average net vertical flux is $\langle B_z\rangle\approx 10$~Gauss. 
The model extends 2.9~Mm below the photosphere up to 32~Mm vertically, with a nonuniform mesh with maximum resolution of 50~km in order to better resolve the low atmosphere up to 15~Mm.
The horizontal extent is $32\times32$ Mm$^2$, solved with a uniform resolution of 62.5~km in both directions.
The second CBP model, {\tt cbp032031}, is obtained from a snapshot of the first one ($t=100$~min) by doubling its resolution in all the three directions and run for half an hour.
This CBP model has roughly the same resolution as the plage models discussed below, in section~\ref{sec:plage}. 
There are several interesting differences in the behavior of chromospheric dynamics and their spectral signatures arising from the change in resolution, but here we will only consider the effects that a change in numerical resolution has on the \oi~line and \mgii~line core statistics.

\begin{figure*}
    \centering
\includegraphics[width=0.9\linewidth]{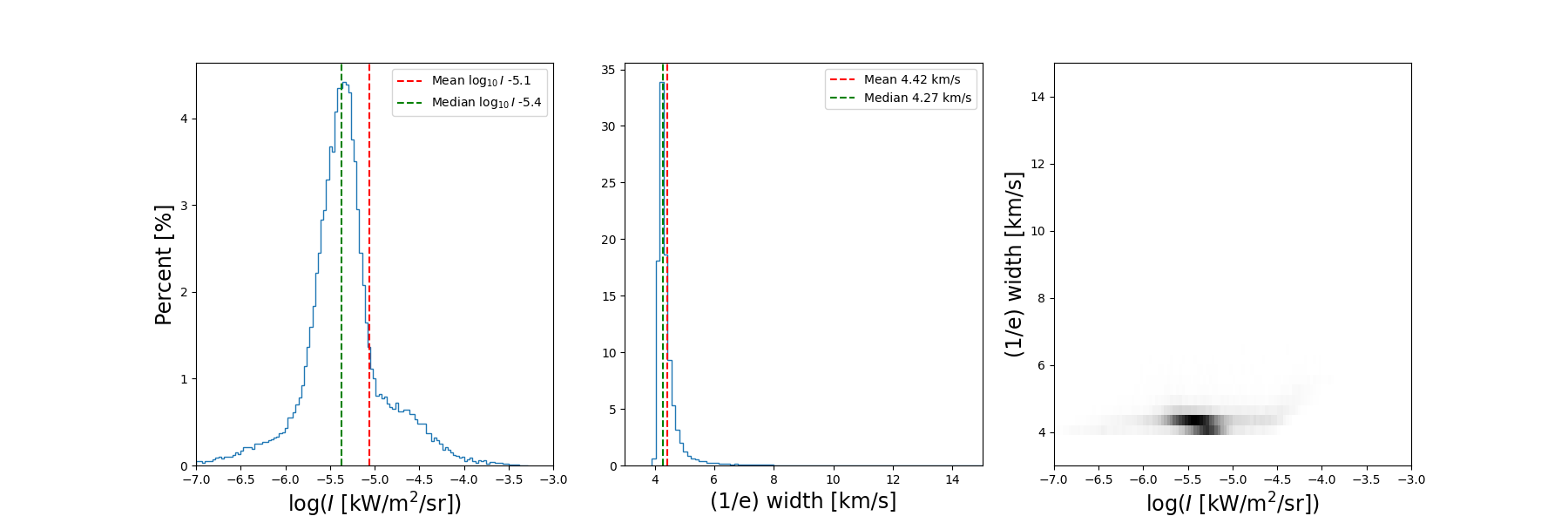}
\includegraphics[width=0.9\linewidth]{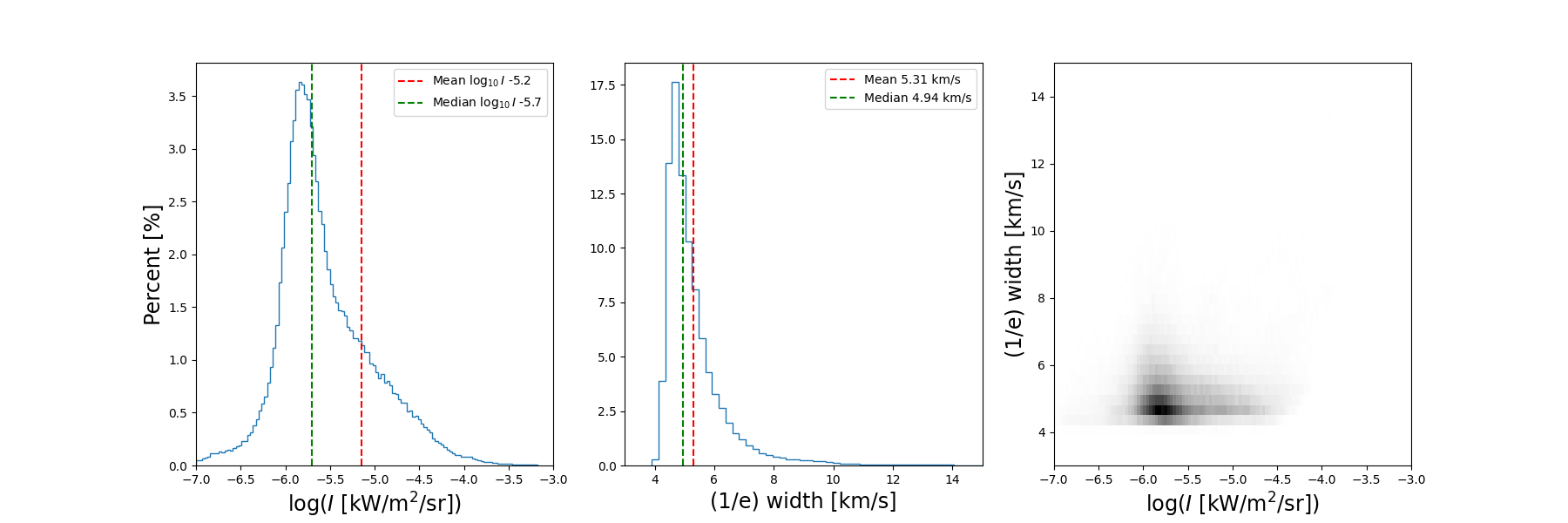}
    \caption{Histograms of the total line \oi~intensities (left panel), $(1/e)$ line widths (central panel), and the JPDF of these quantities for coronal bright point models. The top row shows the lower resolution  model. The higher resolution  model (bottom row) shows larger widths.  The red dashed lines in the left and central panels shows the mean intensities and widths respectively, while the green dashed lines shows the median values.}
    \label{fig:cbp_oi_histo_jpdf}
\end{figure*} 

\begin{figure*}   
    \centering  \includegraphics[width=0.9\linewidth]{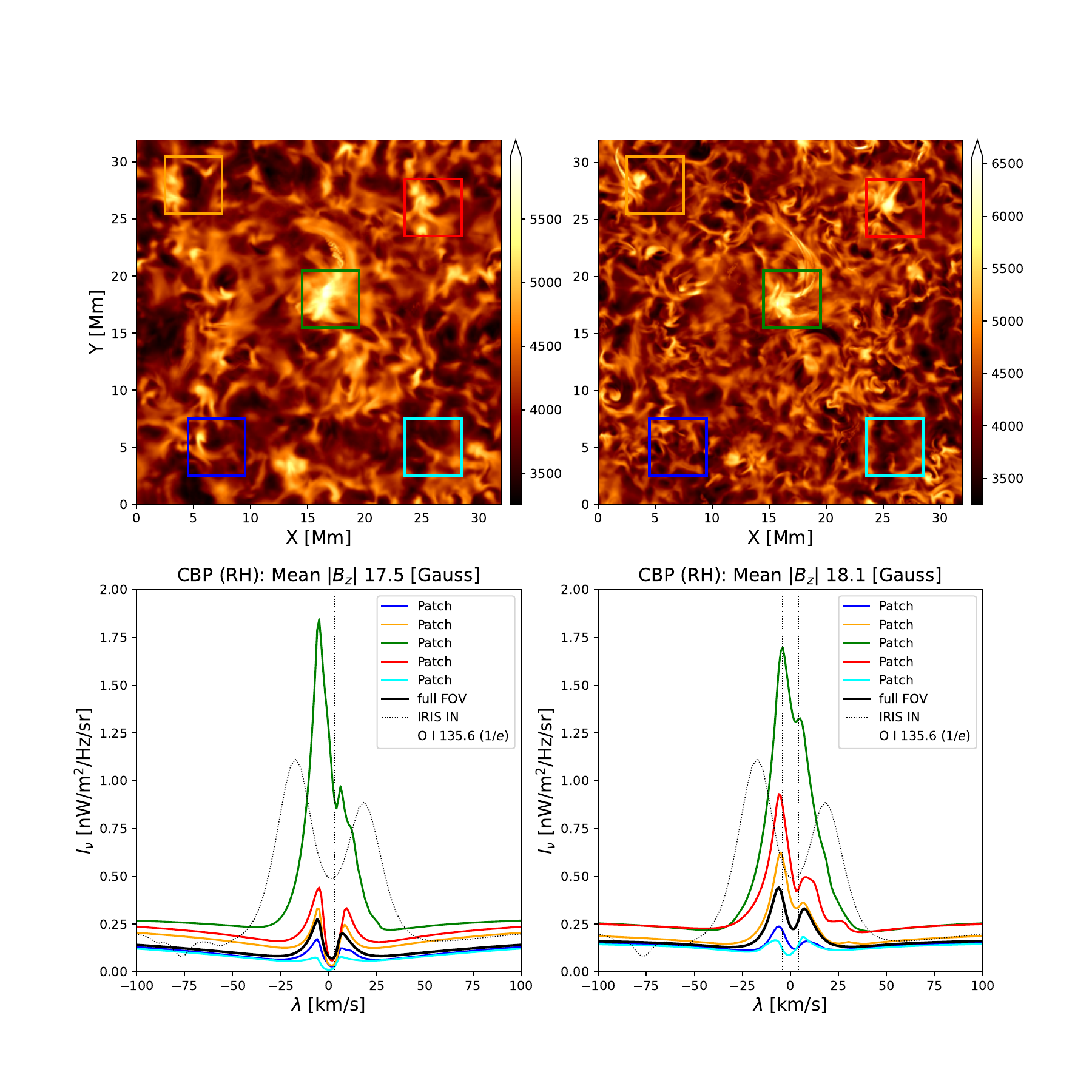}
    \caption{\mgii~core line widths for coronal bright point models. The left column shows \mgii~k line intensities and profiles, for the low resolution model, while the right column shows the same for the high resolution model. Overplotted in the lower row profile plots, in dashed black, is an average IRIS-observed internetwork profile taken with a very large dense raster on February 25, 2014 at 18:59:47 (UTC).}
    \label{fig:cbp_mg_width}
\end{figure*}

The general structure of the CBP vertical field in the photosphere and the emission in the \mgii~k and \oi~lines in the CBP models is shown in Figure~\ref{fig:cbp_model}. Both spectral lines show a brightening above the central parasitic positive polarity near $(X,Y)\approx(16, 18)$~Mm. 
In addition, at the snapshot chosen, a jet-like feature stretching from the center of the model is visible upwards towards the ``north''.
This is due to an erupting chromospheric fibril \citep[see][]{2023ApJ...958L..38N}.
We also find enhanced emission in a ring-like structure surrounding the central polarity. This ring corresponds to the footpoints of the fan surface of the CBP, where magnetic field lines originating from the central polarity return through the chromosphere to the surrounding opposite-polarity regions. A similar feature is found in IRIS data where a region of enhanced intensities in \oi~135.6~nm (also seen in \siiv\ 
and \cli) appears to outline a dome-like structure that separates two flux systems, presented in Figures 6 and 16 of \citet{2023ApJ...959...87C}, occurring at a location where flux has emerged into a pre-existing field configuration. 

The morphology of the map of \oi\ widths is very similar in structure to what was found in the weak field quiet Sun models described in \ref{sec:qs}. One reason for this is that the CBP model is embedded within a magnetic topology meant to simulate a coronal hole, similar to the quiet Sun model with relatively low field strength, but with 
a relatively large flux imbalance in the surrounding areas.
This is borne out by considering Figure~\ref{fig:cbp_oi_histo_jpdf} which shows histograms of the total line intensities and widths of the \oi~line. The average widths are smaller in the low resolution model than in the higher resolution model, by about 1~\kms. A more striking difference is that the high resolution model has a much greater population of line widths larger than 6~\kms\ than the low resolution model. The JPDF of the higher resolution model is similar in shape to what is found in the weakest field quiet Sun model discussed above, and the average and median widths are similar in amplitude to both weaker field quiet Sun models. 

However, the difference between the low and high resolution \oi~widths, and therefore most probable velocity spread, may be insufficient to encourage greater \mgii~core widths  as can be seen in Figure~\ref{fig:cbp_mg_width}: the \mgii~core widths measured are 9.9~\kms\ for the low resolution model and 10.1~\kms for the high resolution model, this difference is smaller than our estimated fitting error (of order 1~\kms) in computing \mgii~core line widths. 

Ignoring for a moment the yellow, red, and green patches which contain regions with stronger magnetic fields, we find that the cyan and blue patches show slightly elevated intensities in the high resolution model. This is also clear when considering the average full FOV profile, which shows a significant increase in intensity compared to the model with lower resolution. The intensity increase is presumably caused more by the increase in vertical resolution, from 50~km to 25~km, than by the horizontal increase from 63~km to 31~km.  We note that the chromospheric scale height is roughly 100~km, and that there is no such intensity increase in the previously discussed quiet Sun models when going from 100~km to 50~km horizontal resolution, when keeping the vertical resolution of 20~km unchanged. In those models the vertical resolution is of the same order as in the high resolution CBP model presented here. 

All the \mgii~core widths in the CBP models fall considerably short of that measured by the IRIS telescope. 
Thus, based on the findings from comparing the relatively weak field {\tt qs072100} and {\tt qs072050} ($\langle|B_z|\rangle\simeq 30$~Gauss) and these CBP models we conclude that increasing the numerical resolution can increase one or both \oi\ line and \mgii\ core line widths, but to a limited extent in weak field regions. 

When considering the yellow, red and green patches associated with stronger magnetic fields we note larger differences in both intensities and widths. This could indicate that higher resolution models will encourage higher turbulent pressures and thereby higher mass loading in the chromosphere, or that the better resolved magnetic field can apply additional lift and higher densities in regions where magnetic fields are significant.

\subsection{Plage model}\label{sec:plage}

Having seen the effects of an increasing average magnetic field in the quiet Sun models, as well as the relatively smaller effects (of order 1~\kms\ for the models considered so far) of numerical resolution, let us turn to plage models where the magnetic field is significantly higher while the magnetic field topology is in some respects somewhat less complex than in the models previously shown. 

Two plage models are considered: {\tt pl024031} and {\tt pl024031\_hion}, which differ in the sense that the latter model is run while solving the non-equilibrium time dependent rate equations for hydrogen (NEQ), while the former is run under the assumption of hydrogen ionization being in LTE, as has been assumed in the previous models described in sections~\ref{sec:qs} and \ref{sec:cbp}. Both plage models have a horizontal resolution of 31~km and a vertical resolution of some 12~km in the photosphere, chromosphere, and transition region. The average $|B_z|$ field strength of these models is much higher than the other models presented, of order $\langle|B_z|\rangle \simeq 200$~Gauss. The flux imbalance is small: $\langle B_z\rangle\simeq-0.15$~Gauss in both models.

\begin{figure*}
\centerline{\includegraphics[width=1.2\linewidth]{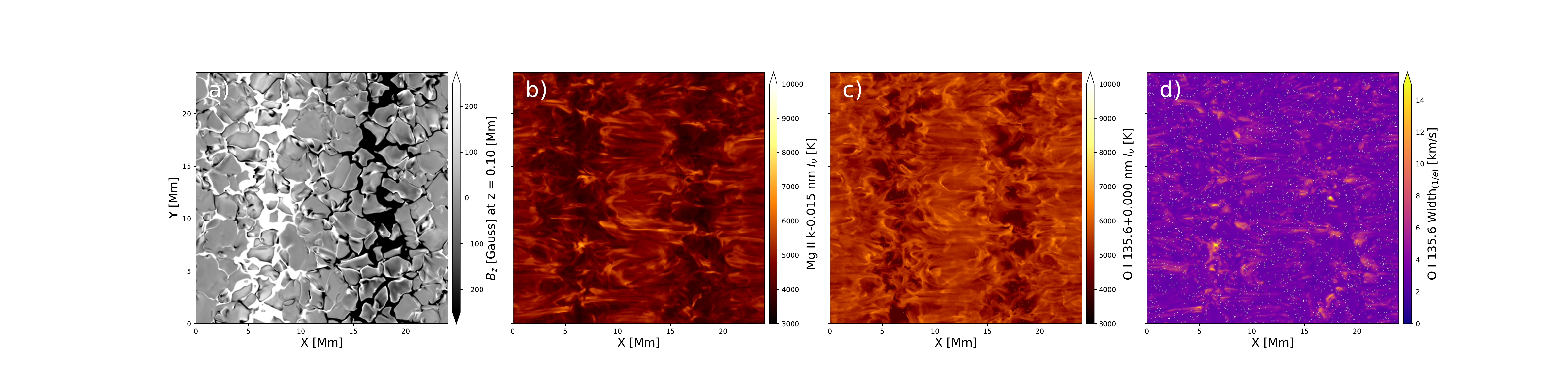}}
\vspace{-0.6cm}
\centerline{\includegraphics[width=1.2\linewidth]{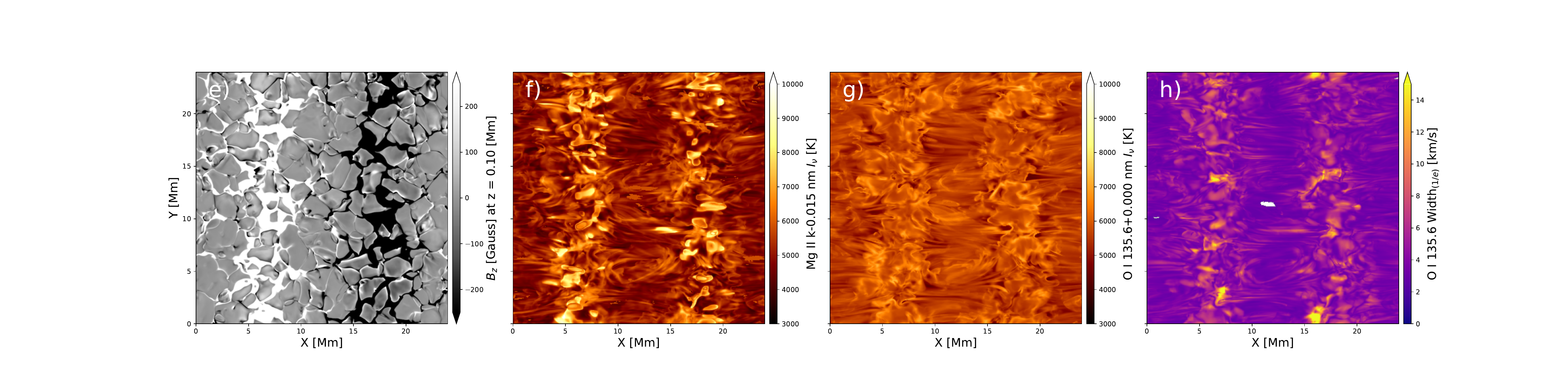}}
    \caption{Plage models. The upper row shows the LTE hydrogen ionization model, while the lower row shows the model with non-equilibrium hydrogen ionization; a) and e): vertical magnetic field at $z=0.1$~Mm, saturated to [-250,250]~Gauss.  b) and f): \mgii~k line intensity at $-0.015$~nm from line center, close to the k$_{2v}$ peak.   c) and g): Line center intensity of the \oi~135.6~nm line. d) and h): \oi~135.6~nm $(1/e)$ widths. }
    \label{fig:plage_model}
\end{figure*}

In Figure~\ref{fig:plage_model} the photospheric magnetic field and emission in the \mgii~k line and \oi~135.6~nm lines are shown for the model  that includes non-equilibrium hydrogen ionization. The strongest magnetic fields are originally confined to two bands of some 3~Mm width with very similar strength, separated by 10~Mm bands of much weaker field. With time these magnetic bands slowly disperse into the weaker field region separating them.  

As discussed in section~\ref{sec:oi_formation} the \mgii~k~$\pm 0.015$~nm and \oi~lines are formed at similar heights and hence their intensities show similar structure: high intensities above the strongest field regions and somewhat weaker emission in between the strong magnetic field bands. The \oi~widths are also greater above the strong field bands and are significantly narrower in the regions in between. Note however that there are large differences between the LTE and NEQ models: While intensities appear slightly depressed above the strongest field regions (for both the \mgii\ and \oi~lines) when LTE is used, intensities are much higher in the non-equilibrium hydrogen ionization model. A similar increase is seen in the \oi~line widths, and the regions above the strongest photospheric fields have significantly greater widths.
The intensity and width characteristics for the NEQ model are similar to what is reported by \citet{2023ApJ...959...87C}
who find that the maps of \oi~line widths on large scales generally resemble the maps of \oi\ intensities. They further find a difference between plage and quiet Sun intensities of roughly a factor 4 which is close to the increase in average widths that we find in the non-equilibrium hydrogen ionization model as discussed below. 

\begin{figure*}
    \centering
\includegraphics[width=0.9\linewidth]{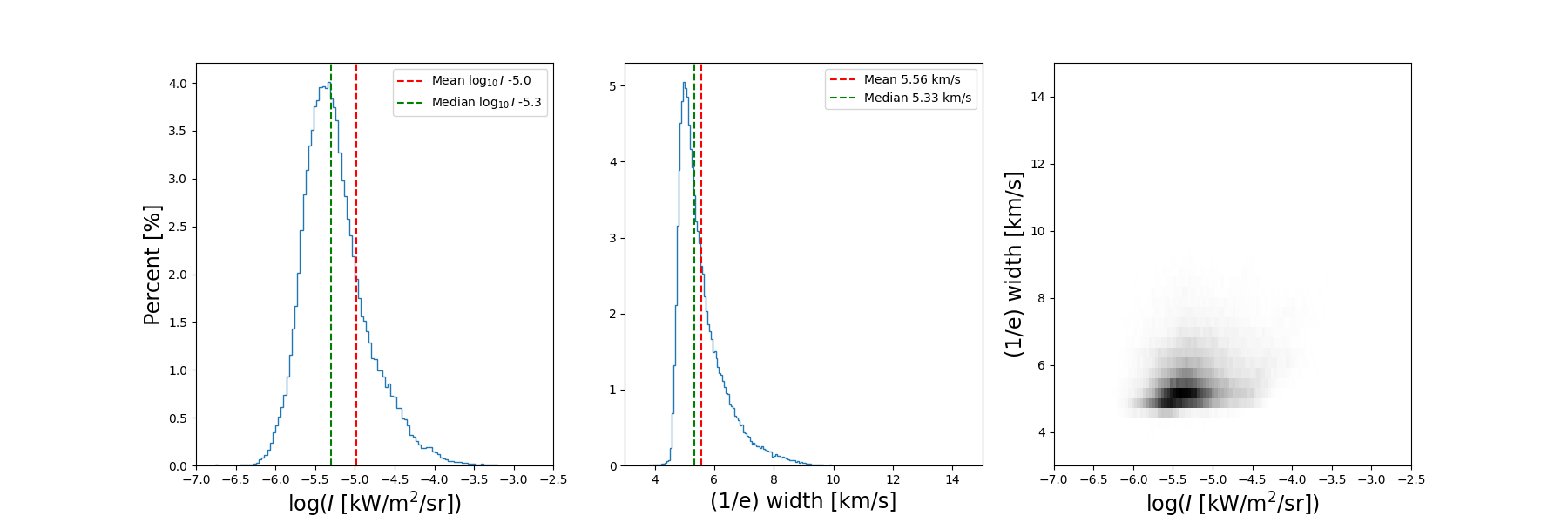}
    \includegraphics[width=0.9\linewidth]{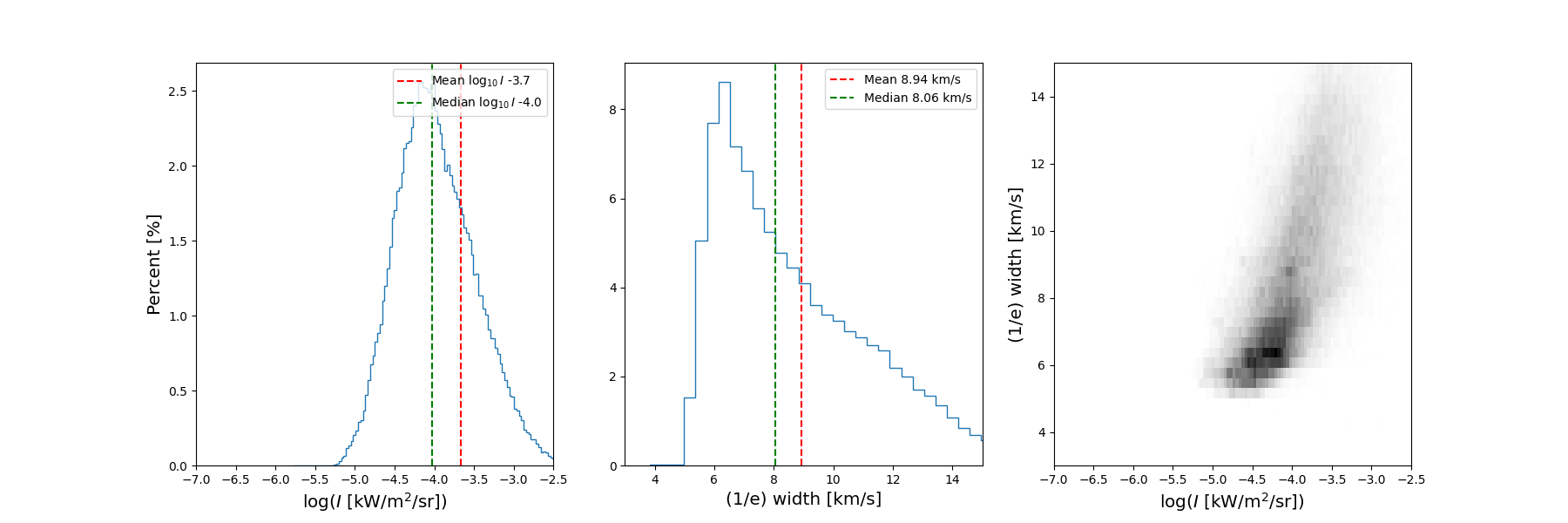}
    \caption{Plage model histograms of the total line \oi~intensities (left panel), $(1/e)$ line widths (central panel), and the JPDF of these quantities for plage models. The top row shows the LTE model, and the bottom row the NEQ hydrogen ionization model. These models are similar in their photospheric vertical magnetic fields $\langle|B_z|\rangle$ which are 217~ Gauss, and 202~Gauss respectively.  The red dashed lines in the left and central panels shows the mean intensities and widths respectively, while the green dashed lines shows the median values.}
    \label{fig:pl_oi_histo_jpdf}
\end{figure*}

Statistical details of the \oi~intensities and widths are shown in Figure~\ref{fig:pl_oi_histo_jpdf}. While the intensities in the equilibrium LTE model are very similar to those found in the quiet Sun models, the non-equilibrium model has mean and median intensities a factor 5 brighter, close to the increase reported from IRIS observations. Likewise, the mean and median \oi~widths are greater in the non-equilibrium model than when LTE is assumed, with median values of $5.33$, and $7.52$~\kms, respectively, and, similarly, mean values of $5.56$, $8.54$~\kms\, respectively. Note that while the LTE model has similar or even smaller widths than the quiet Sun and coronal bright point models, the non-equilibrium hydrogen ionization models have greater widths than what is found in the quiet Sun simulations. We also find a correlation between the \oi~line intensity and line width in the non-equilibrium model. This is in line with the findings from the IRIS observations \citep{2023ApJ...959...87C}.

\begin{figure*}   
    \centering  \includegraphics[width=0.9\linewidth]{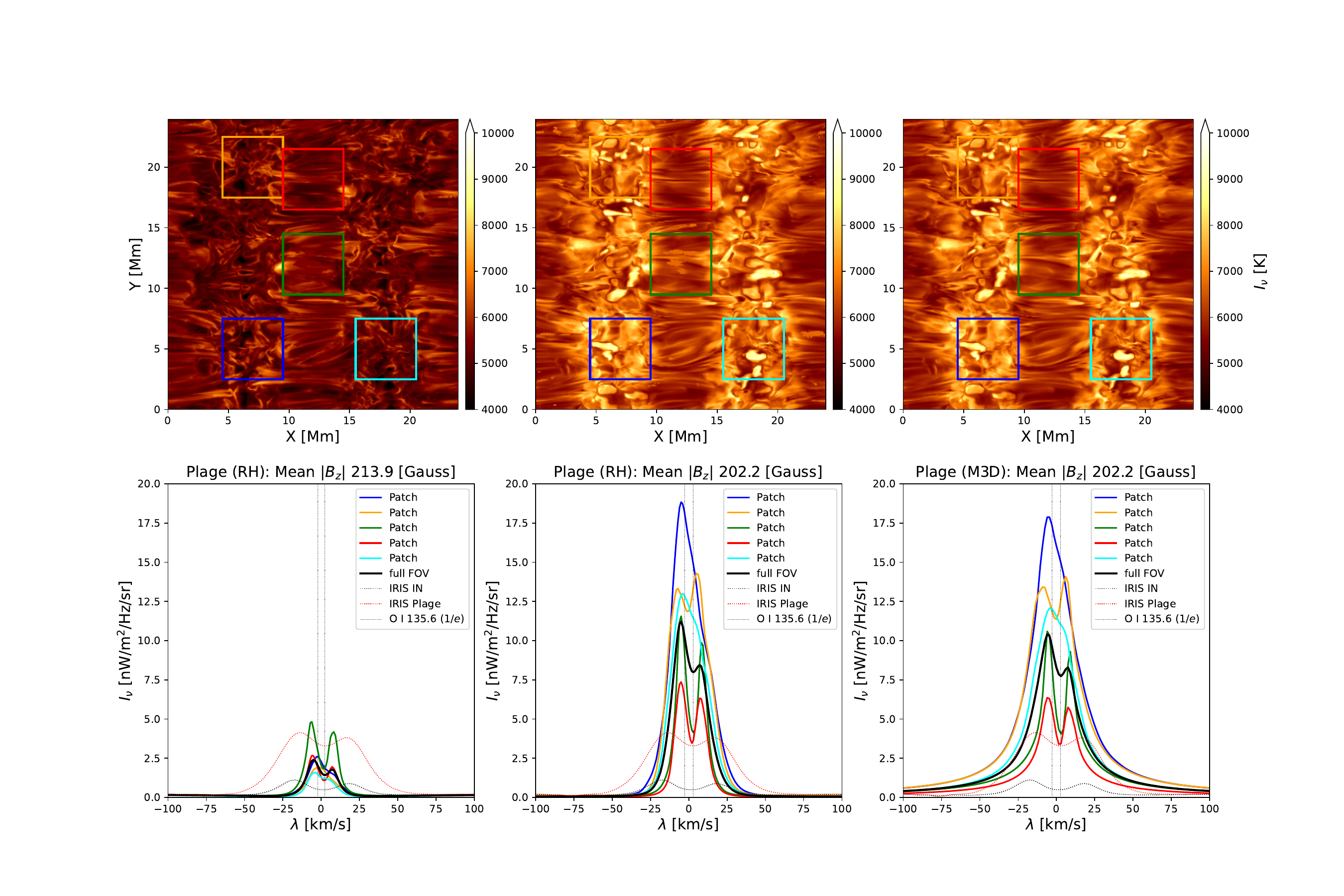}
    \caption{\mgii~core line widths for the LTE (left) and NEQ hydrogen ionization (middle and right) models. The layout is the same as in figure~\ref{fig:qs_mg_width}, but the third column in the figure shows the results when the Multi3D code is used to compute \mgii~line profiles instead of RH1.5D. The red dashed lines show the average line profile of plage measured by IRIS from an observation of NOAA AR 12296 taken March 8 2015 at 15:19:05 (UTC).
    For comparison, the dashed black line is an average IRIS-observed internetwork profile taken with a very large dense raster on February 25, 2014 at 18:59:47 (UTC).}
    \label{fig:pl_mg_width}
\end{figure*}

The resultant \mgii~profiles for the plage simulations are shown in Figure~\ref{fig:pl_mg_width}. The left and central panels of the top row show that for the plage case the difference between LTE and NEQ treatment of hydrogen ionization is quite dramatic: Intensities are much higher in the bands where the magnetic field is strong when non-equilibrium hydrogen ionization is included. This difference is also seen in the \mgii~core line width shown in the lower left and central panels. The LTE plage model has an average \mgii\ core line width of 10.1~\kms, while in the non-equilibrium case 13.0~\kms\ is found. Thus, widths are greater in the NEQ case than when hydrogen populations are assumed to be in LTE. Note that while the intensities are higher than what is typically observed in IRIS plage, the computed \mgii~widths are narrower; in the plage region picked out from the IRIS observation of NOAA AR 12296 on March 8 2015 at 15:19:05 (UTC)\footnote{See Figure~2 of \citet{2023ApJ...944..131H} for \mgii\ core images and the regions chosen to measure average line profiles from this observation.} we find a width of 27~\kms. 

However, for the plage model, there are significant differences in the \mgii\ line profiles calculated with the Multi3D code, as opposed to the RH1.5D code used in the previous examples. This is shown in  the right panels of figure~\ref{fig:pl_mg_width}. While the general character of the map of \mgii~core intensities is very similar when using the two codes, the core widths become wider when accounting for lateral radiative transport. We find a width of 18.1~\kms\ for the full FOV in this case. 
Furthermore, when looking at the individual patches we find widths of 20.1, 21.6, and 20.1~\kms\ for the blue, yellow, and cyan patches that are centered over strong magnetic field regions, while the red, at 16.5~/kms, and green, at 14.5~\kms, patches more resemble modeled quiet sun widths at high resolution and relatively high field strengths. 
% Take a look at the blue and yellow patch cases!
In fact, the widths that are measured above the strong magnetic field areas have values that are close to commensurate with those observed, though we caution that intensities are too high (by a factor $\sim 2$ for the full FOV). In addition, the k$_{2v}$, k$_{2r}$ peak separation in this model is small compared to ``typical'' observed IRIS plage profiles, though we caution that single peaked \mgii\ profiles are also often found in subsets of observed plage.  

\begin{figure*}   
    \centering
    \includegraphics[width=0.8\linewidth]{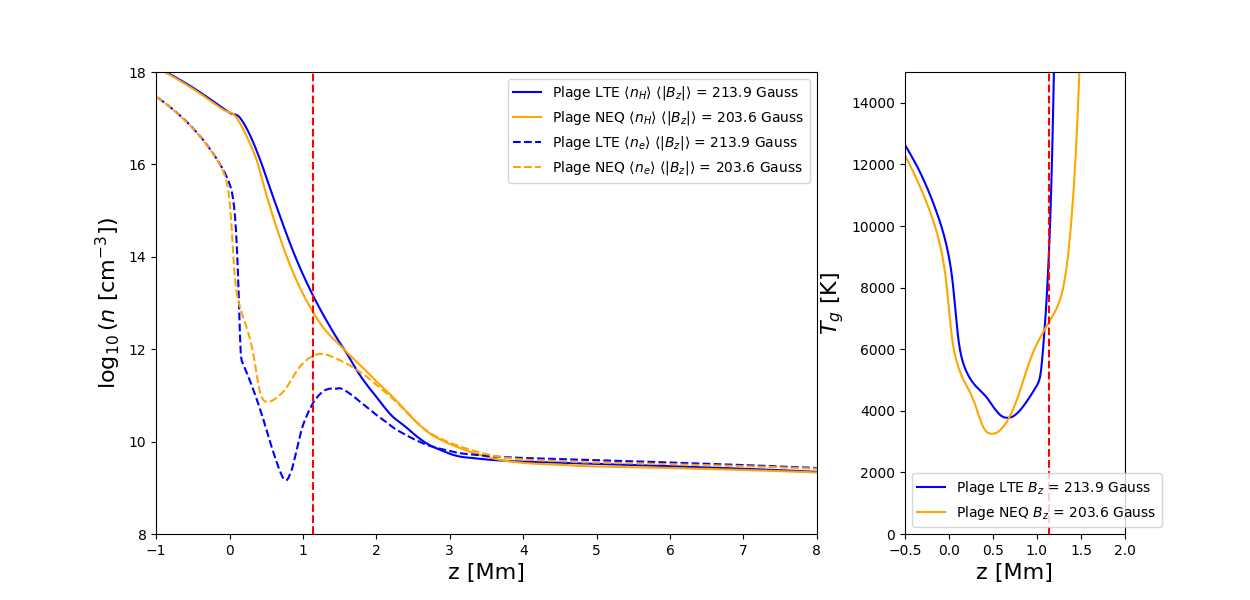}
    \caption{Mean electron ($n_{\rm e}$) and total hydrogen ($n_{\rm H}$) densities as a function of height $z$ for the plage models LTE and NEQ hydrogen ionization. The red dashed lines shows the approximate height of the formation of the \oi~line $z_{\rm fm}$.}
    \label{fig:Plage_n_Tg}
\end{figure*}

Looking into the chromospheric structure of the LTE and NEQ models we find, not surprisingly, that the electron densities are much higher in the NEQ models throughout the chromosphere, as expected \citep[e.g.][]{2007A&A...473..625L} and as shown in Figure~\ref{fig:Plage_n_Tg}. Non-equilibrium ionization also impacts the temperature structure throughout the chromosphere; we find a lower temperature in the lower chromosphere in the NEQ model, and a higher temperature from some 800~km above the photosphere and upwards to the transition region. Interestingly the temperature profile in the NEQ model shows a step in chromospheric temperature with temperatures rising slowly from $6\,500$ to $7\,500$ between 1~Mm and 1.5~Mm. The average height of the transition region is also higher than what is found in the LTE model, but in both models the transition to coronal temperature occurs at a somewhat smaller height, some 1 to 1.5~Mm above the photosphere, than in the quiet Sun models. 

\subsection{Flux emergence model}\label{sec:fe}

Finally, consider a ``flux emergence'' model, {\tt sw072050}, computed assuming LTE hydrogen ionization, in which a sheet of magnetic flux of strength 2000~Gauss, oriented along the $y$-axis is injected for a period of some 2.5~hours at the bottom boundary of a quiet Sun model as described in section~\ref{sec:qs}. At the time of the snapshot considered here the injection at the bottom boundary ceased several hours ago, but the injected field is emerging through the photosphere and into the outer solar atmosphere at the timestep discussed here, and  the average field strength has at this time risen to
$\langle|B_z|\rangle \simeq 100$~Gauss. Thus, the total flux emerging to the photosphere in the computational box is of order $5\times 10^{21}$~Mx, which is typical of an emerging active region \citep[e.g][]{2017ApJ...842....3N}.  In many respects this model is similar to the {\tt nw072100} emerging flux simulation described in \citet{2023ApJ...944..131H}, but is run at double the resolution (50~km vs. 100~km) and has a somewhat larger flux imbalance; $\langle B_z\rangle=6.9$~Gauss vs. $\langle B_z\rangle=0.6$~Gauss.

\begin{figure*}
    \centering
 \centerline{\includegraphics[width=1.2\linewidth]{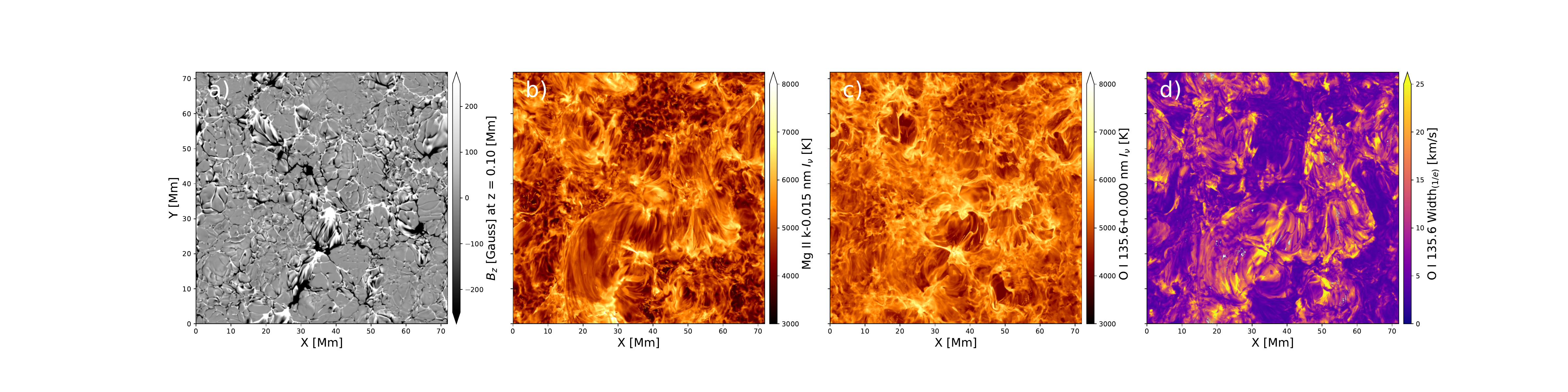}}
    \caption{Flux emergence model. a): vertical magnetic field at $z=0.1$~Mm, saturated to [-250,250]~Gauss.  b): \mgii~k line intensity at $-0.015$~nm from line center, close to the k$_{2v}$ peak.   c): Line center intensity of the \oi~135.6~nm line. d): \oi~135.6~nm $(1/e)$ widths.}
    \label{fig:fe_model}
\end{figure*}

The structure of the solar atmosphere is dramatically different in this simulation than in those previously shown, with large patches of opposite polarity kiloGauss fields interacting in several locations. As seen in Figure~\ref{fig:fe_model} the field has a strong impact on the diagnostics discussed in this paper: The \mgii~core shows heightened emission above magnetically strong regions and especially where opposite polarities are near one another, there are several loops visible, joining these opposite polarity patches. The \oi\ intensities are bright in roughly the same locations as \mgii, and show strongly heightened emission in regions where opposite polarities meet, i.e., regions of active flux emergence and/or cancellation. 

The \oi\ spectral line widths are larger than what is found in the quiet Sun models, but not by much; we find a median of 7.66~\kms and a mean of 9.67~\kms. 

\begin{figure*}   
    \centering  \includegraphics[width=0.6\linewidth]{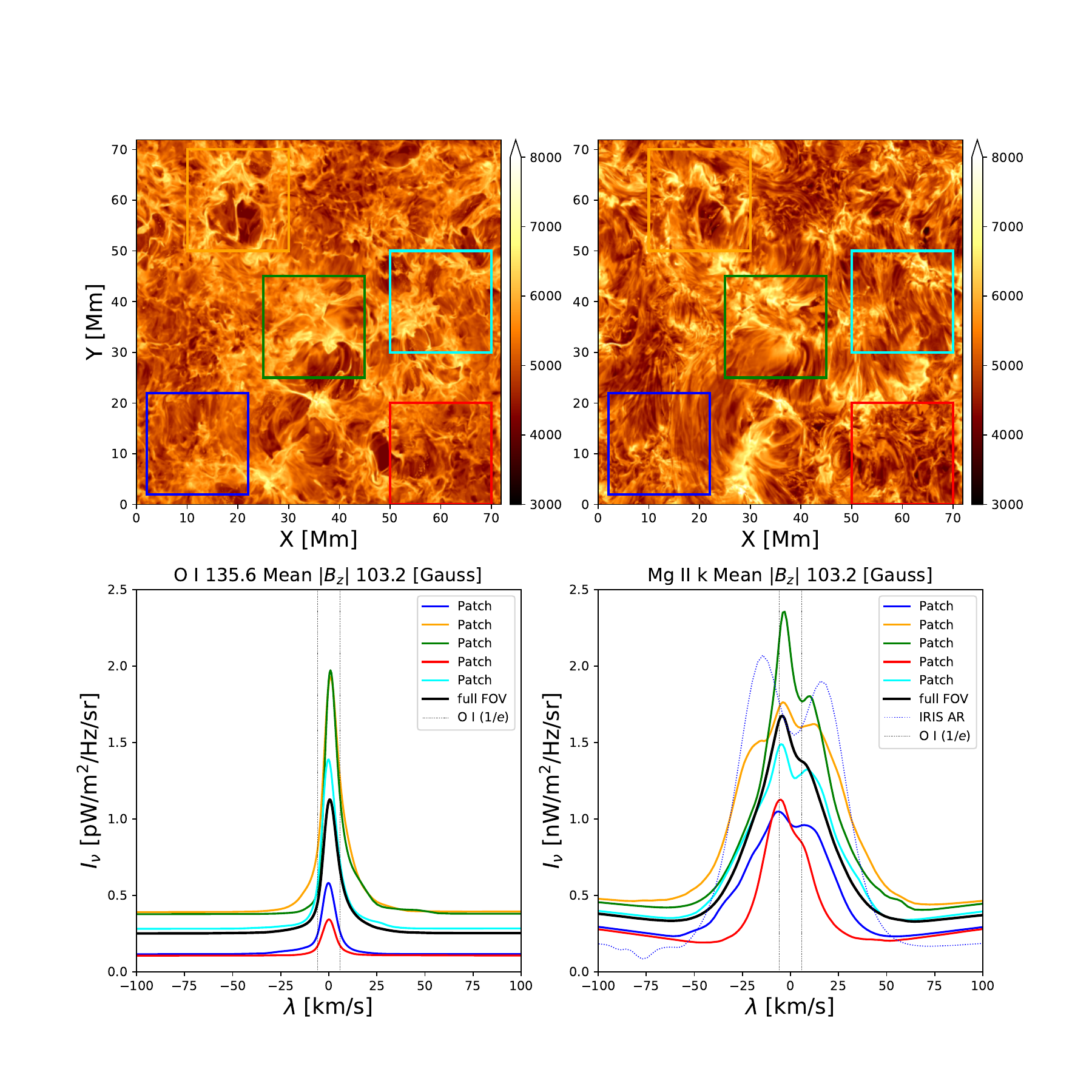}
    \caption{Line core intensities and average spectral profiles for the flux emergence model: The left panels shows the \oi~135.6~nm line, while the right panels the \mgii~k line core. As in previous figures, the profiles are averaged over patches covering $20\times 20$~Mm$^{2}$ while the average profiles for the entire field of view is shown in black. An average active region IRIS spectrum from NOAA AR 12296 taken
    March 8 2015 at 15:19:05 (UTC) is shown in the lower right panel with a dashed black line. }
    \label{fig:fe_ca_oi_mg_width}
\end{figure*}

In the lower middle panel of figure~\ref{fig:fe_ca_oi_mg_width} we show examples of \oi\ spectra averaged over patches of size $20\times 20$~Mm$^2$ the locations of which are shown in the row above as well as the average spectrum for the full field of view. The right panels shows the \mgii~k$_3$ intensity and the \mgii~k line core spectra. These are brighter and broader than what is found in the quiet Sun models, and comparable to what is measured in observed IRIS spectra of an active region; the average active region profile shown is from NOAA AR 12296 taken March 8 2015 at 15:19:05 (UTC). However note that HMI \citep{2012SoPh..275..207S} magnetic field data show this AR experiencing strong flux emergence in the days prior to the IRIS observation, but that this emergence has tapered off by the time the spectra shown were acquired, and the flux emergence rate is slow at the time of the IRIS observation (of order $2\times10^{20}$~Mx$\,$cm$^{-2}$ or less per hour, A.A.~Norton, personal communication, April 30, 2026). It is not clear whether this rate is sufficient to push significant mass into the chromosphere. At any rate, the full FOV \mgii\ core line width is 24.8~\kms\ in this model, which is close to what is observed in the ``typical'' AR 12296.

\begin{figure*}   
    \centering  \includegraphics[width=0.9\linewidth]{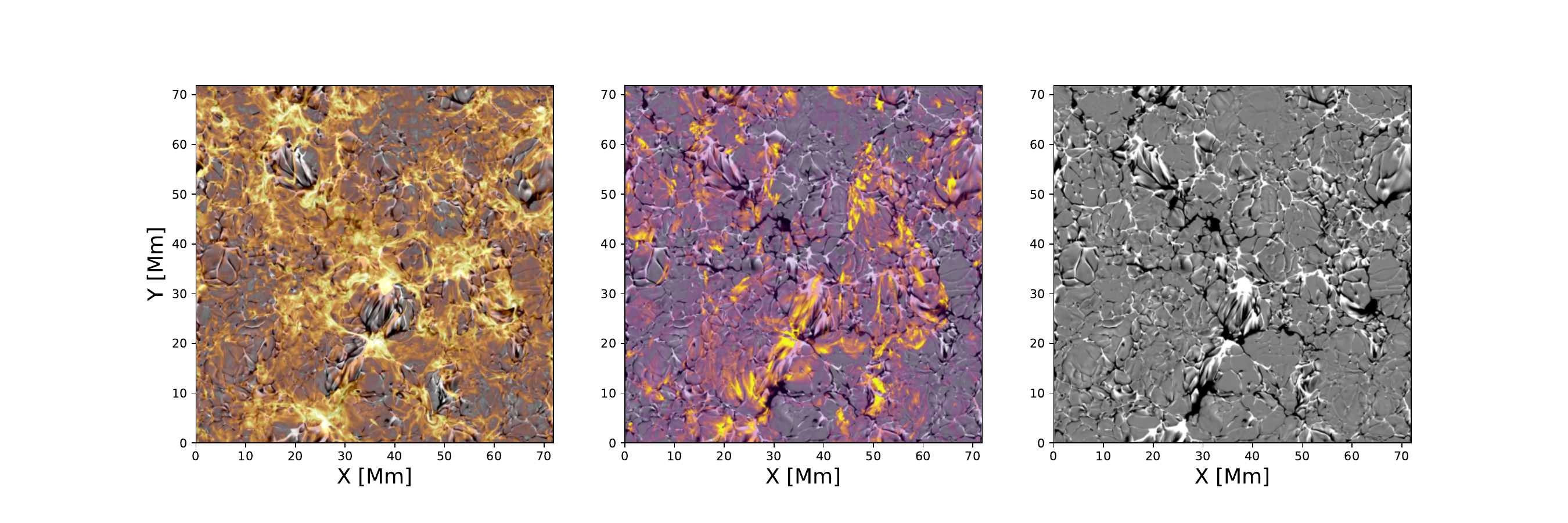}
    \caption{Correlations between sites of flux emergence and cancellation and \oi\ intensities (left panel) and widths (middle panel) in the flux emergence model. The color tables describing the intensities and widths are transparent at low values, but opaque for higher values. These are overplotted the magnetic field $B_z$ at 100~km above the photosphere, which is also shown in the right panel. The field is saturated at $\pm 250$~Gauss in all three panels.}
    \label{fig:fe_oi_itot_oi_wmom}
\end{figure*}

\citet{2023ApJ...959...87C} also point out the correlation between the \oi\ 135.6~nm intensity at locations where flux
concentrations of opposite polarities are in very close proximity, i.e., touching or almost touching, and in regions of flux emergence. Figure~\ref{fig:fe_oi_itot_oi_wmom} shows that the same correlation in the simulated data from this model which has several sites of both flux emergence and cancellation. The figure also shows that large \oi\ widths do not follow the same pattern, sometimes avoiding sites of opposite polarites entirely. This is also in accordance with the findings from that observational study. 

\section{Discussion}

The \oi~135.6~nm spectral line is formed in the upper chromosphere at heights where the magnetic field can be expected to dominate the energetics and dynamics. Additionally, the line is optically thin, and hence opens a unique window into this region of the atmosphere,  providing an unambiguous view of velocities, turbulent non-thermal motions, and electron densities. Specifically, the \oi\ line is formed in essentially the same location as the cores of the brighter and more readily observable \mgii~h\&k lines, giving the possibility of understanding what the properties of the \mgii~profile are telling us about the upper chromosphere. 

It has been clear \citep{2015ApJ...809L..30C} that the observed width of the \mgii\ line is a result of high densities and/or large non-thermal velocities in the chromosphere. Such elevated densities can be provided in several ways; by raising the chromospheric temperature and hence scale height, by the action of a turbulent velocity field, or by the action of the vertical component of the Lorentz force.  The latter can occur as the result of lifting of material into the chromosphere during flux emergence, or from the ambient field providing support. Additionally, non-equilibrium hydrogen ionization will on average increase the chromospheric electron density as well as modify the temperature structure. Finally, horizontal radiative transport will broaden \mgii\ core profiles.

The measurement of \oi\ widths provides the possibility of sorting which of these possible broadening mechanisms is most likely to be active. Both IRIS observations and the models presented here show that the \oi\ widths are lie the range of 5--10~\kms, with an average of 8~\kms\ found in the observations, and somewhat lower in most of the models presented here.  We find that on larger scales \oi\ widths and intensities follow similar spatial patterns, as do \mgii\ and \oi\ core intensities, in both plage and, to a lesser extent, in quiet Sun regions. These patterns are also found in IRIS observations where plage widths are seen to map the \oi\ intensities on large scales. Similarly the low intensity internetwork quiet Sun has very narrow widths, while network regions are reported to show a significant increase in
the line width. 

At the densities expected in the chromosphere, the measured \oi\ widths should, on average, be sufficiently wide to provide a turbulent non-thermal pressure $\rho w_{\rm nt}^2$, that is of similar order, or sometimes even exceeds, the thermal pressure. 
However, in the quiet Sun models discussed in section~\ref{sec:qs} we only find an increase of $\sim 1$~\kms\ in the \oi\ turbulent velocities, not large enough to significantly change the force balance  when the average unsigned vertical magnetic field strength $\langle|B_z|\rangle$ is raised from 17 to 62 Gauss. Since significant, of order 5~\kms, changes in the \mgii\ core width occur as the field strength is raised to $\langle|B_z|\rangle=62$~Gauss this implies that processes other than turbulent pressure are also active in setting the \mgii\ width.  

This point is buttressed by the the strong field plage models described in section~\ref{sec:plage} where the turbulent velocities are of the same order or only slightly higher than what is found in the quiet Sun and CBP models. The widths of the \mgii\ core are much smaller than what is found in observations when considering the model with hydrogen ionization treated in LTE. However, when non-equilibrium hydrogen ionization is included, broader profiles are found, and, when the effects of horizontal radiative transfer are taken into account, these profiles approach measured plage profile widths. As seen in Figure~\ref{fig:Plage_n_Tg}, the density and temperature structure is in this case also set by the much larger electron density in the NEQ plage model. 

While the \mgii\ line widths approach observed values in some scenes and locations, there still remains some space for improvement. In the Quiet Sun models at the ``correct'' average photospheric magnetic field of $\langle|B_z|\rangle\simeq 60$~Gauss the \mgii\ width falls at least 5~\kms\ short of what is typically measured; the weaker field models are another 5~\kms\ narrower. 

The CBP models models discussed in section~\ref{sec:cbp} show the consequences of increasing the numerical spatial resolution in our models. Doubling resolution both vertically and horizontally impacts both the dynamics and diagnostics of the chromosphere. We find that ~\oi\ widths can increase by roughly 1~\kms\ on average while \oi\ intensities remain similar. On the other hand, \mgii\ intensities rise by a factor 2, while the \mgii\ widths remain relatively unchanged when the resolution is improved, except above the strongest field regions where they increase by a few \kms. It should be noted that we do not find a similar increase in \mgii\ intensities when only increasing the horizontal resolution in our quiet Sun models. This is perhaps not surprising considering the 50~km vertical resolution of the low resolution CBP model, which is a fairly large distance compared to a chromospheric scale height of 100~km. The increase in horizontal resolution, to 50~km or better, does give an additional boost, again of order 1~\kms, to the width of the \mgii\ core in the quiet Sun models. 

% intensities QS and CBP

In addition to \mgii~k line widths, core intensities give information on chromospheric heating and structure. The quiet Sun models (section~\ref{sec:qs}) show intensities that are well below those observed, at least for the $\langle|B_z|\rangle=17.1$~Gauss model, increasing somewhat in the
$\langle|B_z|\rangle=27.7$~Gauss model. Low \mgii\ intensities are also found in the coronal bright point models, discussed in section~\ref{sec:cbp}, that have relatively weak magnetic fields ($\langle|B_z|\rangle\simeq 18$~Gauss. It is only when the field is increased to values of $\langle|B_z|\rangle=62$~Gauss in the quiet Sun models that we find intensities approaching those observed.

% intensities Plage and FE

In the plage models we find \mgii\ intensities that are smaller than what is observed in typical plage as measured by IRIS in the LTE hydrogen model, but increasing to 1.5-3 times the observed values when NEQ hydrogen ionization is imposed. The average vertical field strength at $z = 750$~km is  $\langle|B_z|\rangle=160$~Gauss over the full field of view and of order $\langle|B_z|\rangle=240$~Gauss when only averaging over the bands located at $x = [3,10]$ and $x=[16,22]$~Mm. These values are fairly close to what is measured with the ViSP and VIM instruments at the DKIST telescope~\citep{2023ApJ...954L..35D}, but a larger number of plage like simulations, varying the topology and strength of the magnetic field, should be carried out before definitive conclusions can be drawn. 

The flux emergence model presented in section~\ref{sec:fe} which has $\langle|B_z|\rangle=100$~Gauss displays mean \mgii\ intensities of the same order as what is observed as ``typical'' AR values in IRIS data as shown in figure~\ref{fig:fe_ca_oi_mg_width}, but we caution that the rate of emergence is low or uncertain in AR 12296 at the time the IRIS observations were made, so it is unclear whether flux emergence increases chromospheric mass loading in this case. 

Reaching chromospheric densities sufficient to recover observed \mgii\ widths entails either increased nonthermal velocites or increased vertical Lorentz forces through more highly resolved magnetic field topology or continual flux emergence. A higher numerical spatial resolution could conceivably improve these possibilities, but since our modeled \oi\ widths are already close to what is observed, adding 5~\kms\ to the \mgii\ width through turbulent pressure increases or Doppler broadening seems unlikely. An increased magnetic lift resulting from a better numerical resolution or additional flux emergence are a possible routes. However, given the large changes seen in the plage models when non-equilibrium hydrogen ionization and 3D radiative transfer were introduced, the effect of these on the quiet Sun and CBP models should be considered first.

%%%%%%%%%%%%%%%%%%%%%%%%%%%%%%%%%%%%%%%%%%%%%%%%%%%%%%%%%%%%%%

\section{Conclusions}

We have considered a set of numerical models of varying resolution, size, and magnetic topology. This was done to attempt to identify the mechanisms that produce average \mgii~k line core widths in excess of 20~\kms\ in essentially all solar scenes.  To do so we consider the importance of non-thermal motions, observed and modeled to be of order 5--10~\kms, in setting the chromospheric structure and scale height. While non-thermal motions undeniably provide a source of chromospheric mass loading, we find that, for quiet Sun, the average strength of the photospheric magnetic field is the most important parameter in setting the \mgii\ core width to values within 5~\kms\ of observed values. For plage, we find models that include non-equilibrium hydrogen ionization and computing \mgii\ intensities with fully 3D radiative transfer are a promising route to reaching observed values. Generally, a combination of resolution, magnetic field, NEQ, 3D radiative transfer, as well as ambipolar diffusion may be required to fully reproduce the observed line widths. 

We also find that the synthetic \oi\ observations show great similarity with IRIS observations. This includes \oi\ intensities increasing in the vicinity of cancellation and flux emergence sites, the similarity of \oi\ intensities with \mgii\ line core intensities, and the patterns of \oi\ widths with chromospheric network and plage regions.
%May want to expand on this with a few examples -- intensities increased close to cancellation & emergence,  similarity of O I intensities and patterns/maps, similarity of O I line widths, etc.

%%%%%%%%%%%%%%%%%%%%%%%%%%%%%%%%%%%%%%%%%%%%%%%%%%%%%%%%%%%%%%
\begin{acknowledgements}
This research has been supported by NASA contracts NNG09FA40C (IRIS) and 80NSSC24K0258. Additional support was provided by the Research Council of Norway (RCN) through its Centres of Excellence scheme, project number 262622.  Furthermore, support  was provided by the European Research Council through the Synergy Grant number 810218 (``The Whole Sun'', ERC-2018-SyG).
The simulations have been run on the Pleiades cluster through the computing project s1061, and s2278 from the High End Computing (HEC) division of NASA and through computational resources provided by Sigma2, the National Infrastructure for High Performance Computing and Data Storage in Norway. The authors also acknowledge the computer resources at the MareNostrum
supercomputing installation and the technical support provided by the
Barcelona Supercomputing Center (BSC, RES-AECT-2021-1-0023,
RES-AECT-2022-2-0002). IRIS is a NASA small explorer mission developed and operated by LMSAL with mission operations executed at NASA Ames Research Center and major contributions to downlink communications funded by ESA and the Norwegian Space Agency (NOSA). 

A.A.~Norton is thanked for providing animations showing magnetograms and the flux emergence rate as a function of time in the hours surrounding the IRIS observation of AR 12296.

\end{acknowledgements}

%%%%%%%%%%%%%%%%%%%%%%%%%%%%%%%%%%%%%%%%%%%%%%%%%%%%%%%%%%%%%%
% WARNING
% Please note that we have included the references below in
% order to compile the document, but we ask you to:
%
% - use BibTeX with the regular commands:
\bibliographystyle{aa} % style aa.bst
\bibliography{oi.bib} % your references Yourfile.bib
% - join the .bib files when you upload your source files
%%%%%%%%%%%%%%%%%%%%%%%%%%%%%%%%%%%%%%%%%%%%%%%%%%%%%%%%%%%%%%

\FloatBarrier %\usepackage{placeins}
\clearpage

\end{document}